\documentclass[twocolumn]{aastex63}

\usepackage{fullpage,amsmath,longtable,graphicx,natbib}
\usepackage{hyperref}

\newlength\tindent
\renewcommand{\indent}{\hspace*{\tindent}}

\begin{document}

\title{Spectroscopic Follow-Up of Discoveries from the NEOWISE Proper Motion Survey}

\author{Jennifer J. Greco}
\affil{Department of Physics and Astronomy, The University of Toledo, Toledo, OH 43606 USA. Jennifer.Greco@rockets.utoledo.edu}

\author{Adam C. Schneider}
\affil{School of Earth and Space Exploration, Arizona State University, Tempe, AZ, 85282 USA}

\author{Michael C. Cushing}
\affil{Department of Physics and Astronomy, The University of Toledo, Toledo, OH 43606 USA. Jennifer.Greco@rockets.utoledo.edu}

\author{J. Davy Kirkpatrick}
\affil{IPAC, MS 100-22, Caltech, 1200 East California Blvd., Pasadena, CA 91125, USA}

\author{Adam J. Burgasser}
\affil{Center for Astrophysics and Space Science, University of California San Diego, La Jolla, CA 92093, USA}



\begin{abstract}

We present low-resolution near-infrared spectra of discoveries from an all-sky proper motion search conducted using multi-epoch data from the {\it Wide-field Infrared Survey Explorer}. Using the data from NEOWISE, along with the AllWISE catalog, \cite{Schneider2016} conducted an all-sky proper motion survey to search for nearby objects with high proper motions. Here, we present a follow-up spectroscopic survey of 65 of their discoveries, which focused primarily on potentially nearby objects (d $<$ 25 pc), candidate late-type brown dwarfs ($\geq$L7), and subdwarf candidates. We found 31 new M dwarfs, 18 new L dwarfs, and 11 new T dwarfs. Of these, 13 are subdwarfs, including one new sdL1 and two new sdL7s. Eleven of these discovereies, with spectral types ranging from M7 to T7 (including one subdwarf) are predicted to be within 25 pc, adding to the number of known objects in the solar neighborhood. We also discovered three new early-type T subdwarf candidates, one sdT1, one sdT2, and one sdT3, which would increase the number of known early-type T subdwarfs from two to five.

\end{abstract}


\section{Introduction}

The census of stars and brown dwarfs in the solar neighborhood expanded dramatically with the launch of the \textit{Wide-field Infrared Survey Explorer} (\textit{WISE}; \citealt{Wright2010WISE}). Discoveries include the third and fourth closest systems to the Sun in WISE J104915.57$-$531906.1AB (hereafter WISE 1049$-$5319AB; \citealt{Luhman2013}) and WISE J085510.83$-$071442.5 (hereafter WISE 0855$-$0714; \citealt{LuhmanW0855}), a substantial increase in the number of known late-type T dwarfs (e.g \citealt{Mace2013}), a new spectral class (Y dwarfs; \citealt{Cushing2011}; \citealt{Kirkpatrick2012}), and the M9 dwarf, WISE J072003.20$-$084651.2, located in the Galactic Plane (hereafter WISE 0720$-$0846; \citealt{Scholz2014}; \citealt{Kirkpatrick2014}).WISE 0720$-$0846, which is an M9.5/T5 binary \citep{Burgasser2015} at a distance of $6.72 \pm 0.05$ pc \citep{Henry2018}, is of particular interest because it passed within 0.25$_{-0.07}^{+0.11}$ pc of the Sun 70 ky ago, in the closest known flyby of a star to the solar system \citep{Mamajek2015}. 


\textit{WISE} was built to survey the entire sky simultaneously in four mid-infrared bands whose central wavelengths are at 3.4 $\mu$m ($W1$), 4.6 $\mu$m ($W2$), 12 $\mu$m ($W3$), \& 22 $\mu$m ($W4$). The four-band cryogenic mission surveyed the sky 1.2 times between January 2010 and August 2010. After the cryogen in the outer tank was depleted, a three-band cryogenic survey was conducted using the $W1$, $W2$, and $W3$ bands, covering an additional 30\% of the sky, until the cryogen in the inner tank was also exhausted in September 2010. Following this, a two-band survey was conducted using only the $W1$ and $W2$ bands \citep{Mainzer2011}. The result of these surveys was two full maps of the sky and 20\% of a third, separated by $\sim$6 months. The data from all of these surveys were combined and used to generate the AllWISE source catalog \citep{Cutri2014}. The \textit{WISE} satellite was then put into hibernation until December 2013, when it was reactivated to search for potentially hazardous near-Earth objects, using the $W1$ and $W2$ bands alone (NEOWISE; \citealt{Mainzer2014}). 

The multi-epoch nature of the \textit{WISE} observations meant that for the first time, all-sky proper motion surveys at infrared wavelengths were possible. \cite{Luhman2014}, \cite{Kirkpatrick2014}, and \cite{Kirkpatrick2016} used the data from the original \textit{WISE} mission with a time baseline of $\sim$6 months, to perform the first all-sky mid-infrared proper motion searches, finding 762, 3525, and 1039 new discoveries, respectively. \cite{Schneider2016} used the NEOWISE data in combination with the AllWISE source catalog to conduct a proper motion survey with a time baseline of $\sim$4 years. The longer time baseline of their survey enabled them to detect significantly more objects at fainter magnitudes than the surveys of \cite{Luhman2014} and \cite{Kirkpatrick2014} (see Figure 8 of \citealt{Schneider2016}). 

The \cite{Schneider2016} survey discovered 20,551 motion objects, of which 1006 were new discoveries. In this paper, we present follow-up observations of 65 of these new discoveries. In \S2, we describe how we selected our targets for follow-up observations. In \S3, we detail the follow-up observations that were conducted and present all of our follow-up spectra. In \S4, we present spectral types and distance estimates for each of our objects. In \S5, we discuss our follow-up observations in detail.

\section{Target Selection}
In order to prioritize follow-up spectroscopic observations, \cite{Schneider2016} identified 128 objects that fell into at least one of three categories: 1) potential late-type brown dwarfs (spectral type $\geq$L7), 2) potential nearby objects (d $<$ 25 pc), and 3) potential subdwarfs (i.e. low metallicity dwarfs). To begin their candidate selection, \cite{Schneider2016} first estimated the spectral types of their new discoveries using available near- and mid-infrared photometry and the k-nearest neighbors method against a training set of objects with known spectral types (see Appendix A of \citealt{Schneider2016} for details). They identified a total of 39 candidates with estimated spectral types later than or equal to L7 and presented spectroscopy of six of these. Distances to all new discoveries were then computed using the photometric-based spectral types, $W2$ magnitudes, and the absolute magnitude-spectral type relations of \cite{DupuyLiu2012}. They identified a total of 46 objects with distance estimates less than or equal to 25 pc and presented spectroscopy of three of these. Finally, a total of 58 potential subdwarfs were identified using both a color cut and a reduced proper motion diagram, and spectroscopy of six of these were presented. 

Here we present near-infrared spectroscopy of 65 additional objects. Of these, 53 were selected from the 128 sources selected by \cite{Schneider2016}: 23 candidate late-type brown dwarfs, 21 potentially nearby objects; and 21 subdwarf candidates. Eleven of these were candidates in more than one category, including WISE J032309.12$-$590751.0 and WISE J101944.62$-$391151.6, which were candidates in all three categories. Three additional objects, WISE J111320.39+501010.5, WISE J121231.97$-$050750.7, \& WISE J145747.55$-$094719.3, were identified as subdwarf candidates early on in the survey based on their high tangential velocities (v$_\text{tan} > 100$ km/s). During gaps in our Right Ascension coverage, we supplemented our target list with additional mid L candidates, observing a total of 7 additional objects. Finally, on one night with particularly poor weather, we observed two bright M dwarf candidates. 
 
\section{Observations}

A summary of all follow-up observations is provided in Table 1. Included in this table are the AllWISE designation for each object (hereafter these will be abbreviated as WISE HHMM $-$ DDMM), the UT date of the observation, the telescope/instrument used to conduct the observations, the total exposure time used for each spectrum, the signal-to-noise of the resultant spectra calculated at the peak intensity in the $J$-band, and the A0 V star observed for calibration purposes. All spectra are plotted in Figures 1 -- 6, sorted by spectral type.  

\startlongtable
\begin{deluxetable*}{cccccc}
\tablewidth{0pt}
\tablecaption{Summary of Observations}
\tabletypesize{\small}
\label{table:sdobservations}
\tablehead{
AllWISE Designation\tablenotemark{a} & UT Date & Telescope/Instrument & Total Exp Time(s) & A0 V Star & S/N\tablenotemark{b}}
\startdata
J000430.66$-$260402.3 & 2016 Aug 3 & IRTF/SpeX & 2151 & HD 225200 & 45 \\
J000458.47$-$133655.1 & 2016 Sep 22 & IRTF/SpeX & 2151 & HD 1154 & 18 \\
J000536.63$-$263311.8 & 2016 Sep 22 & IRTF/SpeX & 2151 & HD 222332 & 19 \\
J000856.39$-$281321.7 & 2016 Sep 21 & IRTF/SpeX & 2151 & HD 225200 & 18 \\
J010134.83+033616.0 & 2016 Sep 22 & IRTF/SpeX & 1434 & HD 6457 & 81 \\
J010631.20$-$231415.1 & 2016 Sep 21 & IRTF/SpeX & 2151 & HD 13433 & 12 \\
J011049.18$+$192000.1 & 2016 Oct 24 & IRTF/SpeX & 1434 & HD 6457 & 46 \\
J013525.38+020518.2 & 2016 Aug 6 & IRTF/SpeX & 2151 & HD 1154 & 22 \\
J022721.93+235654.3 & 2016 Aug 3 & IRTF/SpeX & 2151 & HD 13869 & 38 \\
J030119.39$-$231921.1 & 2016 Aug 3 & IRTF/SpeX & 1912 & HD 19622 & 28 \\
J030919.70$-$501614.2 & 2016 Jul 18 & {\it Magellan}/FIRE & 1374 & HD 8811 & 43 \\
J031627.79+265027.5 & 2016 Aug 6 & IRTF/SpeX & 2151 & HD 19600 & 35 \\
J032309.12$-$590751.0 & 2016 Jul 18 & {\it Magellan}/FIRE & 1374 & HD 325 & 64 \\
J032838.73+015517.7 & 2016 Aug 6 & IRTF/SpeX & 2151 & HD 18571 & 24 \\
J033346.88+385152.6 & 2016 Sep 21 & IRTF/SpeX & 1673 & HD 21038 & 51 \\
J034409.71+013641.5 & 2016 Sep 21 & IRTF/SpeX & 2151 & HD 21686 & 14 \\
J034858.75$-$562017.8 & 2016 Jul 18 & {\it Magellan}/FIRE & 1099 & HD 325 & 24 \\
J041353.96$-$202320.3 & 2017 Jan 16 & IRTF/SpeX & 1673 & HD 25754 & 25 \\
J041743.13+241506.3 & 2016 Feb 24 & IRTF/SpeX & 2151 & HD 25175 & 62 \\
J053424.45+165255.0 & 2016 Feb 24 & IRTF/SpeX & 1434 & HD 35036 & 97 \\
J054455.54+063940.3 & 2016 Sep 21 & IRTF/SpeX & 1434 & HD 35153 & 158 \\
J061429.77+383337.5 & 2016 Feb 24 & IRTF/SpeX & 717 & HD 45105 & 270 \\
J062858.69+345249.2 & 2016 Feb 24 & IRTF/SpeX & 1434 & HD 45105 & 48 \\
J063552.52+514820.4 & 2017 Nov 22 & IRTF/SpeX &  714 & HD 45105 & 16 \\
J084254.56$-$061023.7 & 2016 Feb 24 & IRTF/SpeX & 2151 & HD 63714 & 69 \\
J085039.11$-$022154.3 & 2016 Feb 24 & IRTF/SpeX & 1434 & HD 79108 & 89 \\
J085633.87$-$181546.6 & 2016 Mar 28 & IRTF/SpeX & 1434 & HD 82724 & 45 \\
J092453.76+072306.0 & 2016 Feb 24 & IRTF/SpeX & 1434 & HD 79108 & 67 \\
J094812.21$-$290329.5 & 2016 Feb 24 & IRTF/SpeX & 1434 & HD 94741 & 60 \\
J095230.79$-$282842.2 & 2016 Feb 24 & IRTF/SpeX & 1434 & HD 81694 & 96 \\
J101944.62$-$391151.6 & 2016 Dec 09 & CTIO/ARCoIRIS & 2880 & HD 89213 & 19 \\
J103534.63$-$071148.2 & 2016 Mar 28 & IRTF/SpeX & 2151 & HD 93346 & 35 \\
J111320.39+501010.5 & 2016 Mar 28 & IRTF/SpeX & 1434 & HD 99966 & 85 \\
J112158.76+004412.3 & 2016 Feb 24 & IRTF/SpeX & 1434 & HD 97585 & 43 \\
J112859.45+511016.8 & 2016 Mar 28 & IRTF/SpeX & 1434 & HD 99966 & 42 \\
J120751.17+302808.9 & 2016 Feb 24 & IRTF/SpeX & 1434 & HD 105388 & 121 \\
J121231.97$-$050750.7 & 2016 Mar 28 & IRTF/SpeX & 1434 & HD 109309 & 117 \\
J121914.75+081027.0 & 2016 Feb 24 & IRTF/SpeX & 1434 & HD 116960 & 60 \\
J122042.20+620528.3 & 2016 Jun 20 & IRTF/SpeX & 1434 & HD 148968 & 44 \\
J123513.87$-$045146.5 & 2016 Jun 26 & IRTF/SpeX & 2151 & HD 109309 & 41 \\
J124516.66+601607.5 & 2016 Feb 24 & IRTF/SpeX & 1434 & HD 118214 & 79 \\
J133520.09$-$070849.3 & 2016 May 10 & IRTF/SpeX & 1434 & HD 122749 & 19 \\
J134359.71+634213.1 & 2016 May 10 & IRTF/SpeX & 1434 & HD 118214 & 31 \\
J143942.79$-$110045.4 & 2016 Feb 24 & IRTF/SpeX & 1673 & HD 136831 & 66 \\
J144056.64$-$222517.8 & 2016 Jun 20 & IRTF/SpeX & 1434 & HD 133466 & 106 \\
J145645.54$-$103343.5 & 2016 Mar 28 & IRTF/SpeX & 1434 & HD 132072 & 78 \\
J145747.55$-$094719.3 & 2016 Mar 28 & IRTF/SpeX & 1434 & HD 132072 & 56 \\
J155225.22+095155.5 & 2016 Jun 20 & IRTF/SpeX & 1434 & HD 136831 & 70 \\
J165057.66$-$221616.8 & 2016 May 10 & IRTF/SpeX & 717 & HD 155379 & 291 \\
J171059.52$-$180108.7 & 2016 May 10 & IRTF/SpeX & 717 & HD 154921 & 279 \\
J171105.08$-$275531.7 & 2016 May 10 & IRTF/SpeX & 717 & HD 157918 & 220 \\
J171454.88+064349.8 & 2016 Mar 28 & IRTF/SpeX & 1912 & HD 161289 & 42 \\
J173551.56$-$820900.3 & 2016 Jul 18 & {\it Magellan}/FIRE & 1374 & HD 131912 & 89 \\
J180839.55+070021.7 & 2016 May 10 & IRTF/SpeX & 2151 & HD 167163 & 32 \\
J182010.20+202125.8 & 2016 Oct 24 & IRTF/SpeX & 1434 & HD 171623 & 8 \\
J183654.10$-$135926.2 & 2016 Oct 24 & IRTF/SpeX & 1075 & HD 172904 & 13 \\
J191011.03+563429.3 & 2016 Jun 20 & IRTF/SpeX & 717 & HD 172728 & 317 \\
J201252.78+124633.3 & 2016 Sep 22 & IRTF/SpeX & 1195 & HD 191082 & 533 \\
J211157.84$-$521111.3 & 2016 Jul 18 & {\it Magellan}/FIRE & 1374 & HD 200523 & 63 \\
J215550.34$-$195428.4 & 2016 Oct 14 & IRTF/SpeX & 1912 & HD 203893 & 10 \\
J221737.41$-$355242.7 & 2016 Oct 24 & IRTF/SpeX & 2151 & HD 202941 & 13 \\
J223444.44$-$230916.1 & 2016 Oct 14 & IRTF/SpeX & 2151 & HD 212643 & 7 \\
J224931.10$-$162759.6 & 2016 Oct 14 & IRTF/SpeX & 2151 & HD 212643 & 24 \\
J230743.63+052037.3 & 2016 Oct 24 & IRTF/SpeX & 1434 & HD 219833 & 54 \\
J234404.85$-$250042.2 & 2016 Sep 22 & IRTF/SpeX & 1434 & HD 225200 & 81 \\
\enddata
\tablenotetext{a}{The prefix for AllWISE sources is WISEA. So for example J000430.66$-$260402.3 should be listed as WISEA J000430.66$-$260402.3.}
\tablenotetext{b}{Calculated at the peak intensity in the $J$-band.}
\end{deluxetable*}

\begin{figure*}
\centerline{\hbox{\includegraphics[angle=90]{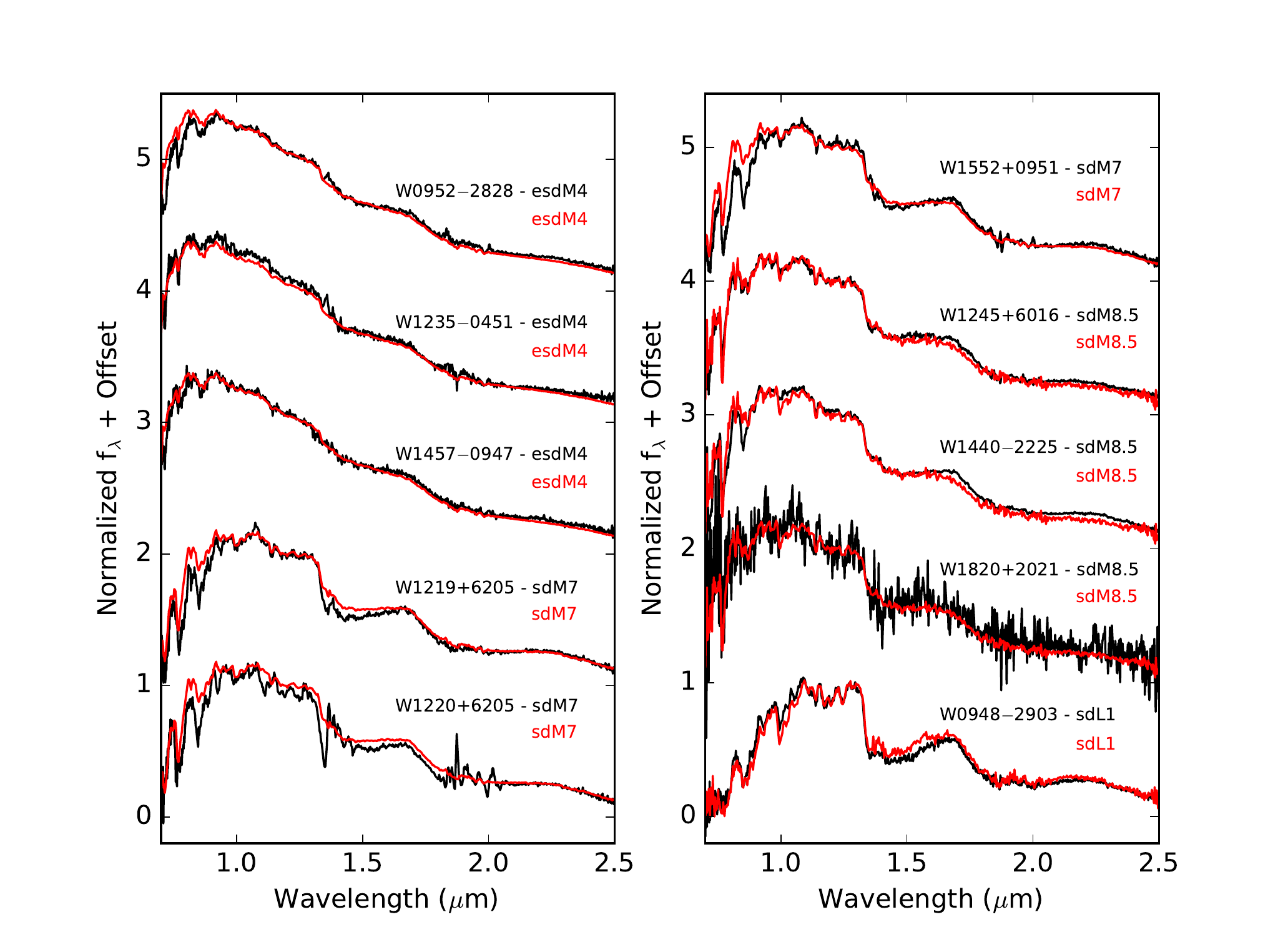}}}
\caption{Spectra of all observed objects, plotted against the appropriate spectral standards. 
\label{fig:spectra1}}
\end{figure*}

\begin{figure*}
\centerline{\hbox{\includegraphics[angle=90]{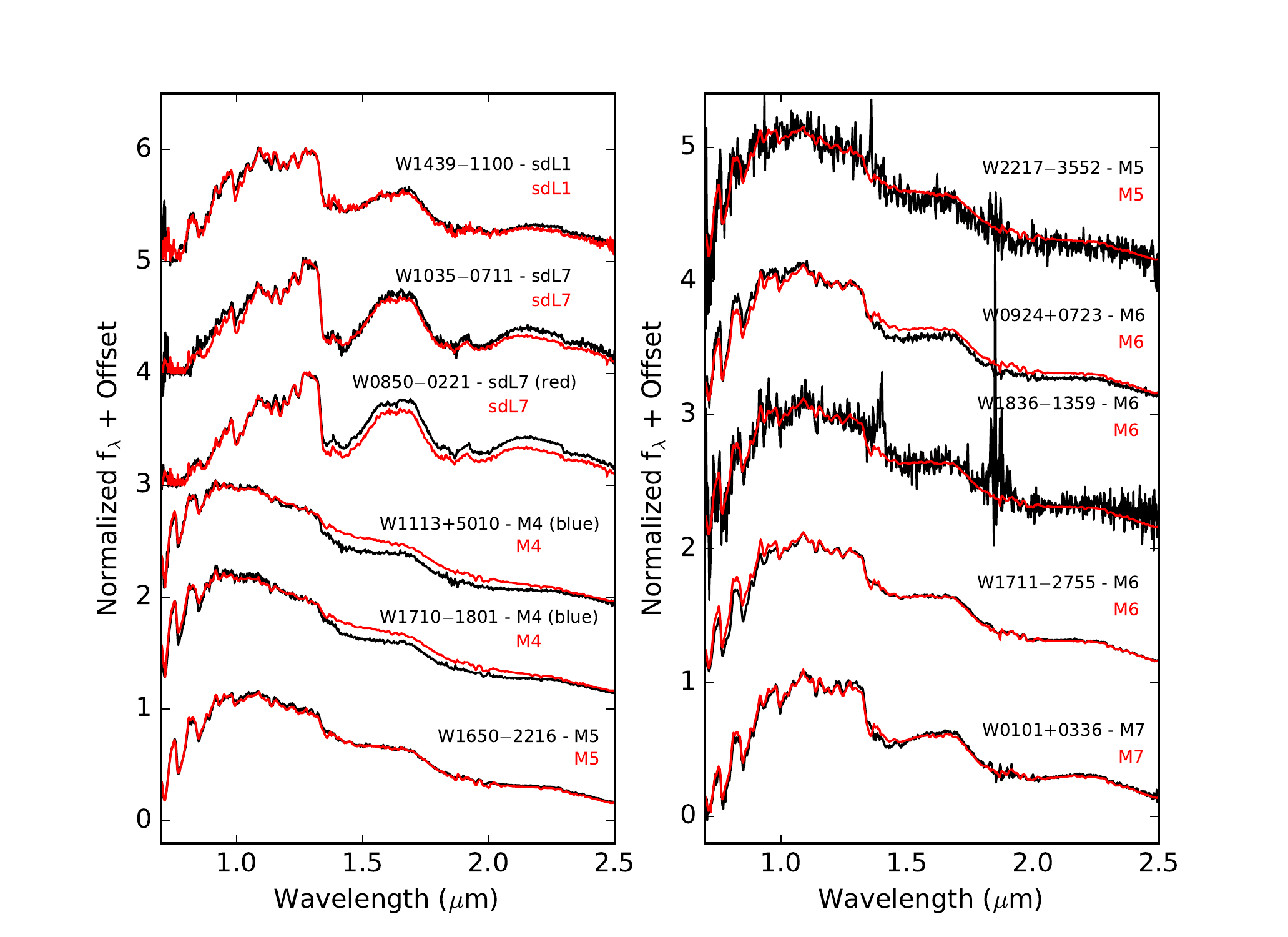}}}
\caption{Spectra of all observed objects, plotted against the appropriate spectral standards. 
\label{fig:spectra2}}
\end{figure*}

\begin{figure*}
\centerline{\hbox{\includegraphics[angle=90]{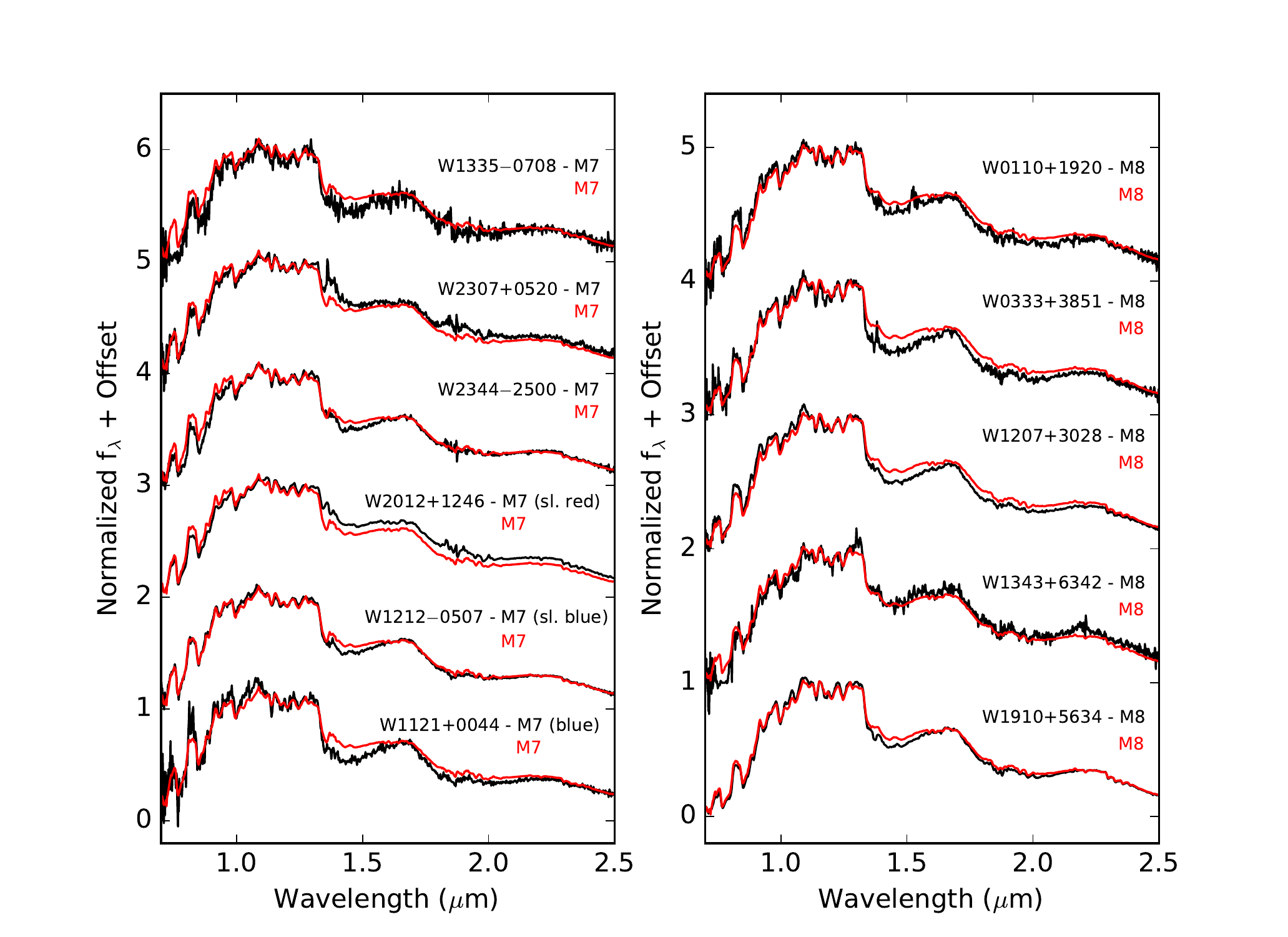}}}
\caption{Spectra of all observed objects, plotted against the appropriate spectral standards. 
\label{fig:spectra3}}
\end{figure*}

\begin{figure*}
\centerline{\hbox{\includegraphics[angle=90]{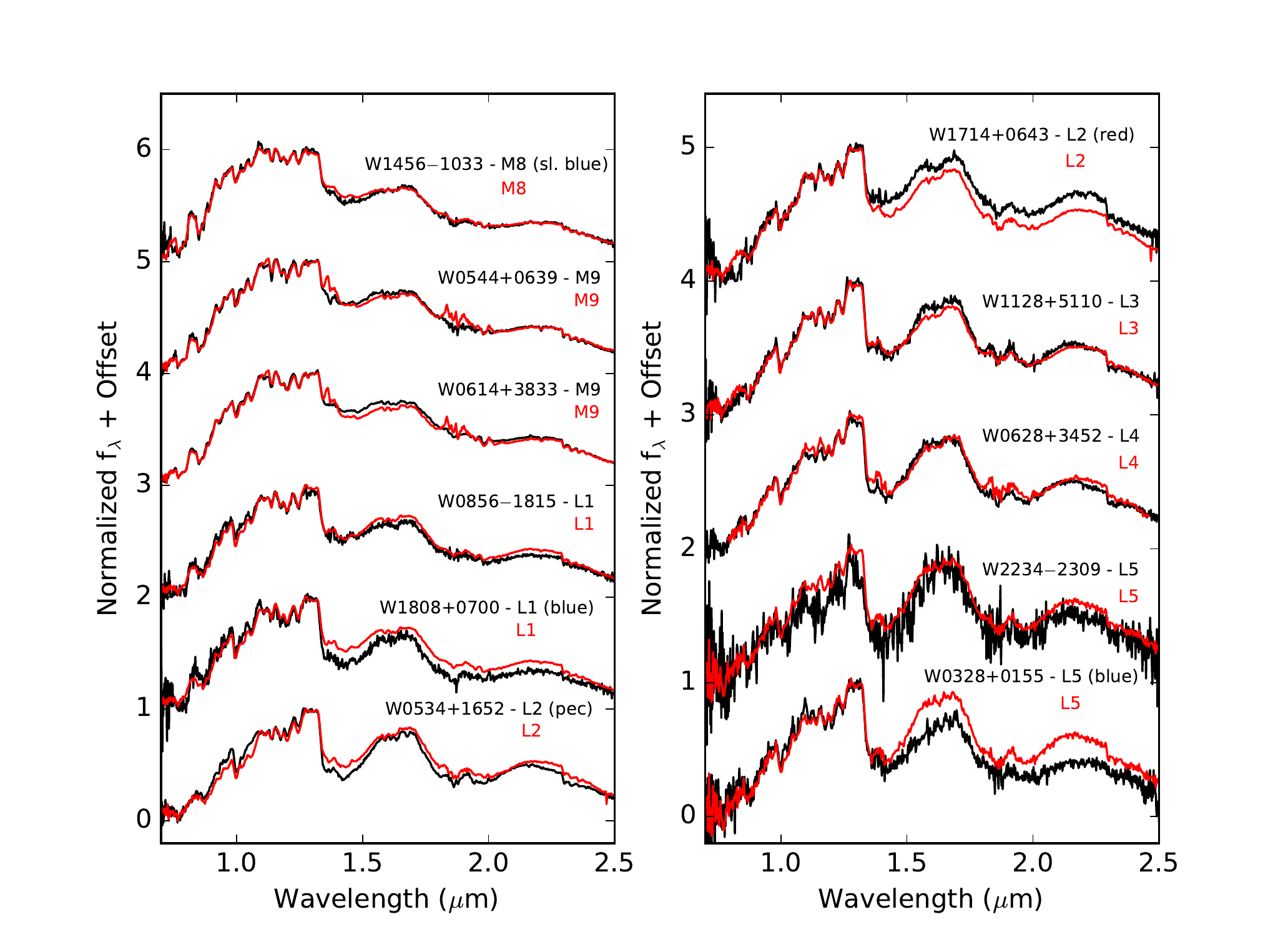}}}
\caption{Spectra of all observed objects, plotted against the appropriate spectral standards. 
\label{fig:spectra4}}
\end{figure*}

\begin{figure*}
\centerline{\hbox{\includegraphics[angle=90]{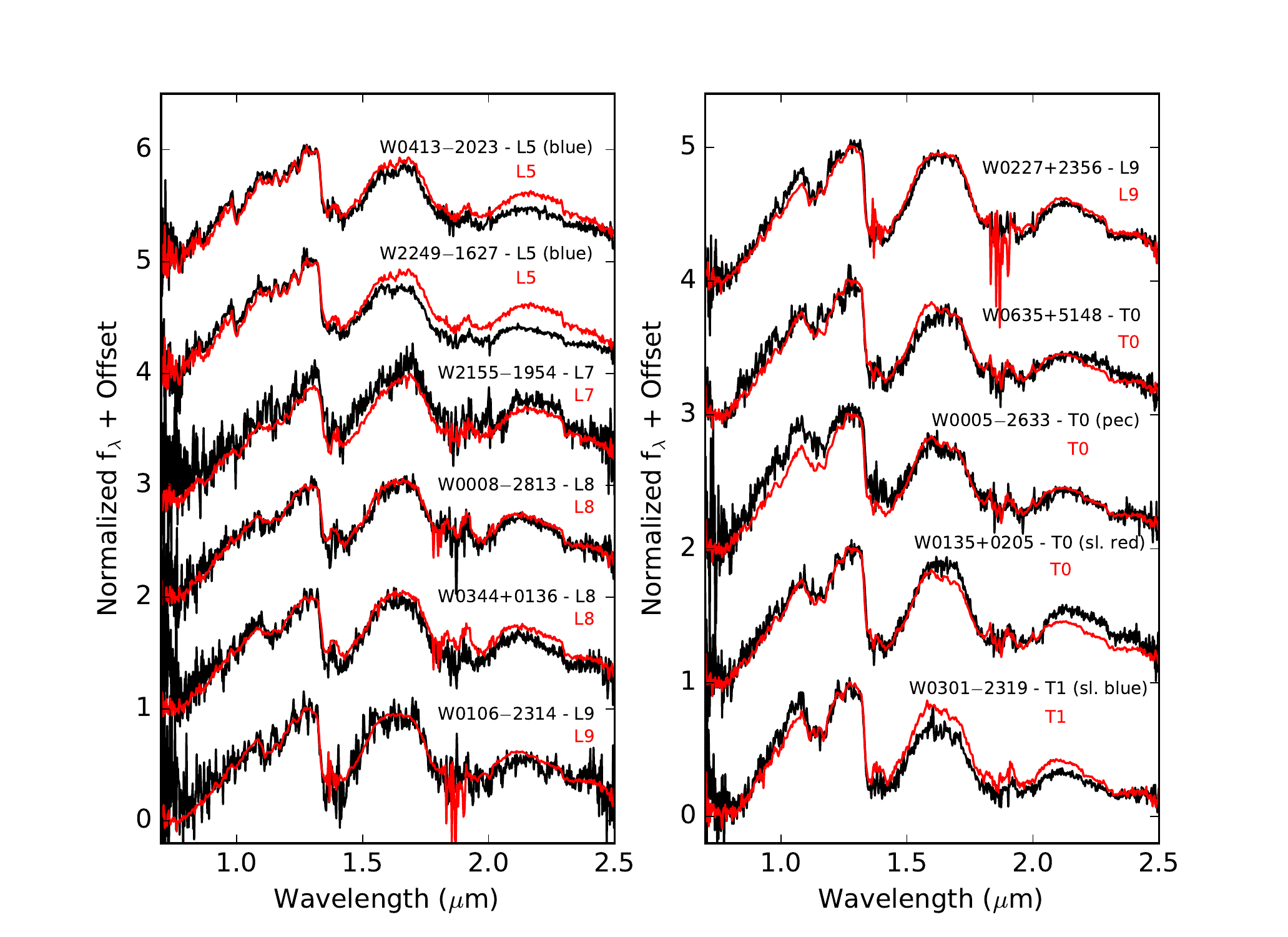}}}
\caption{Spectra of all observed objects, plotted against the appropriate spectral standards. 
\label{fig:spectra5}}
\end{figure*}

\begin{figure*}
\centerline{\hbox{\includegraphics[angle=90]{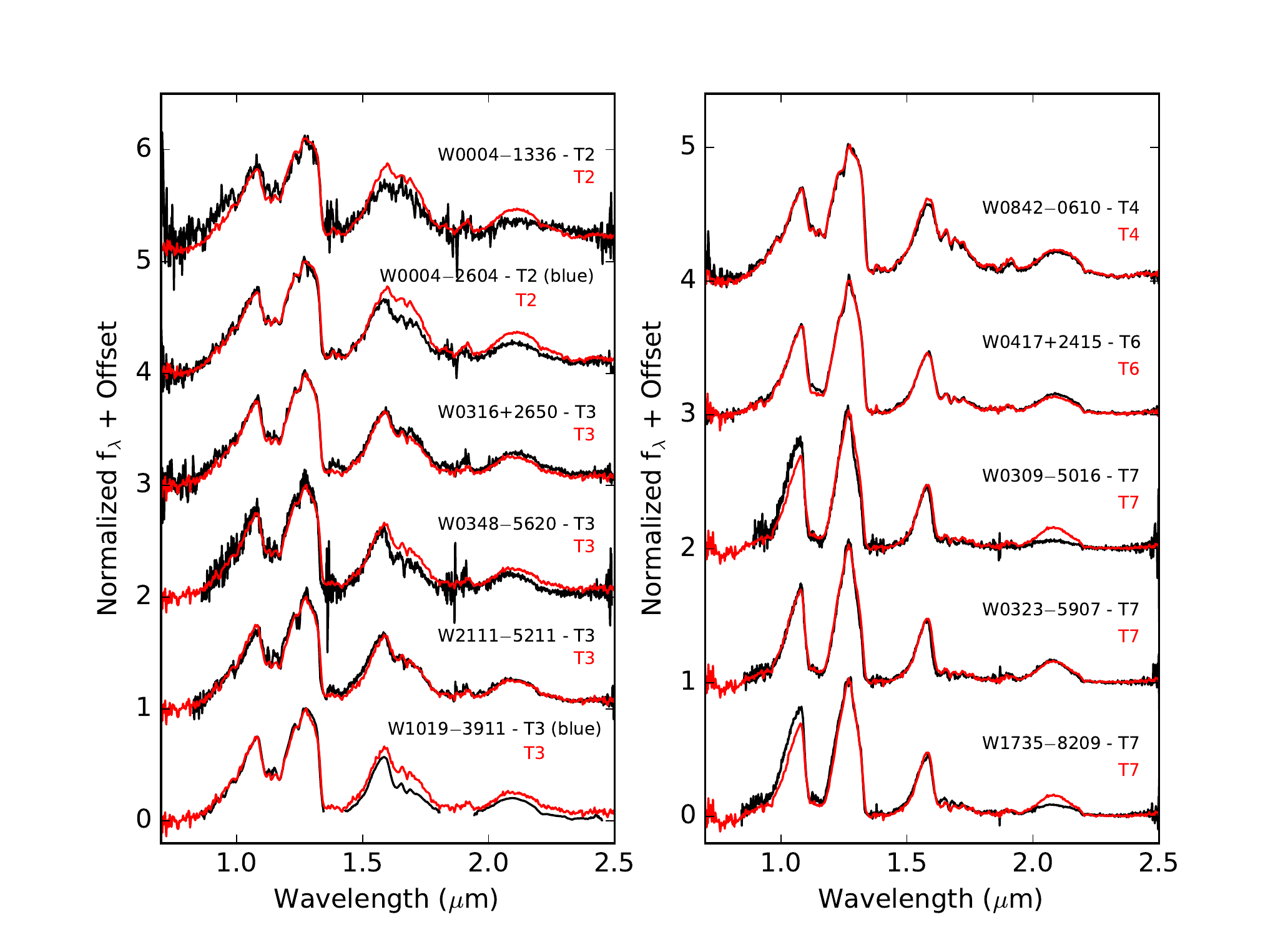}}}
\caption{Spectra of all observed objects, plotted against the appropriate spectral standards. 
\label{fig:spectra6}}
\end{figure*}

\subsection{IRTF/SpeX}

Spectra of 59 objects were obtained using the SpeX spectrograph \citep{Rayner2003Spex} on the NASA Infrared Telescope Facility (IRTF) on Mauna Kea. Observations were conducted between the dates of UT 24 Feb 2016 and UT 22 Nov 2017 (see Table 1 for full list of observation dates). The data were collected in prism mode spanning a wavelength range of 0.8--2.5 $\mu$m with a resolution of $R\equiv \lambda/\bigtriangleup \lambda = 250$, using either the $0\farcs5$-wide slit or the $0\farcs3$-wide slit aligned to the parallactic angle. For each object, a series of exposures were taken using an ABBA nod pattern along the 15$\arcsec$ long slit. Additionally, an A0 V star was observed at a similar airmass to each object and used for telluric correction and flux calibration. The data were all reduced using the Spextool package (\citealt{Vacca2003Spex}; \citealt{Cushing2004Spex}). 

In order to spectral type our subdwarf candidates, we require spectra of subdwarf standards. One of us (A.J.B) obtained spectra of 16 M and L subdwarf standards using IRTF/SpeX. Observations were conducted between the dates of UT Sep 17 2003 and UT Dec 21 2006. Data were collected in prism mode, as discussed above, and reduced using the Spextool package (\citealt{Vacca2003Spex}; \citealt{Cushing2004Spex}). A list of these standards, their spectral types, the references for those spectral types, and the details of those observations are listed in Table 2. Spectra of these objects are shown in Figure 7.

\begin{deluxetable*}{llllllll}
\tabletypesize{\scriptsize}
\tablewidth{0pt}
\tablecolumns{8}
\tablecaption{Subdwarf Standards}
\label{table:sdobservations}
\tablehead{
 \colhead{Object} \vspace{-0.25cm} & \colhead{Other} & \colhead{Optical} & \colhead{Spectral Type} & \colhead{UT Date} &  \colhead{Exp Time}\tablenotemark{a} & \colhead{A0 V Star}\tablenotemark{a} & \colhead{S/N\tablenotemark{b}}\\
 \colhead{Name} \vspace{-0.005cm} & \colhead{Designation} & \colhead{Spectral Type} & \colhead{Reference} & & \colhead{(s)} &  & }
\startdata
LP 51$-$133 & LHS 217 & esdM0\tablenotemark{c} & \cite{Kirkpatrick2010Stds} & 2006 Dec 21 &  320 & HD 33654 & 412 \\
LP 857$-$48 & LHS 375 & esdM4 & \cite{Gizis1997} & 2005 Mar 23 & 540 & HD 125299 & 318 \\
LP 589$-$7 & $\cdots$ & esdM5 & \cite{GizisReid1999} & 2004 Sep 05 & 1080 & HD 13936 & 318 \\
LP 258$-$28 & LHS 2023 & esdM6 & \cite{ReidGizis2005} & 2005 Mar 23 & 1080 & HD 58729 & 318 \\
APMPM J0559$-$29037 & $\cdots$ & esdM7 & \cite{Schweitzer1999} & 2005 Dec 31 & 1440 & HD 41473 & 159 \\
LEHPM 2$-$59 &  $\cdots$ & esdM8 & \cite{Burgasser2006} & 2004 Sep 09 & 720 & HD 32855  & 120 \\
LP 625$-$2 & LHS 3181 & sdM2 & \cite{Raiz2008}\tablenotemark{d} & 2004 Jul 25 & 720 & HD 143459 & 758 \\
LP 803$-$27 & LHS 407 & sdM5 & \cite{Gizis1997} & 2004 Jul 25 & 480 & HD 133772 & 639 \\
LP 645$-$78 & LHS 1074 & sdM6 & \cite{ReidGizis2005} & 2004 Sep 08 & 1080 & HD 18735 & 173 \\
LP 440$-$52 & LHS 377 & sdM7 & \cite{Gizis1997} & 2004 Mar 12 & 480 & HD131951 & 514 \\
2MASS J01423153+0523285 & $\cdots$ & sdM8.5\tablenotemark{e} & \cite{Burgasser2007} & 2003 Sep 17 & 720 & HD 18571 & \phn60 \\
SSSPM J1013$-$1356 & $\cdots$ & sdM9.5 & \cite{Scholz2004} & 2004 Mar 12 & 720 & HD 88025 & 165 \\
2MASS J17561080+2815238 & $\cdots$ & sdL1\tablenotemark{f} & \cite{Kirkpatrick2010Stds} & 2005 Oct 20  & 960 & HD 160557 & \phn78 \\
SDSS J125637.13$-$022452.4 & $\cdots$ & sdL3.5 & \cite{Burgasser2009} & 2005 Mar 23  & 1080 & HD 111744 & \phn87 \\
2MASS J16262034+3925190 & $\cdots$ & sdL4 & \cite{Burgasser2007} & 2004 Jul 23 & 480 & HD 153345 & 429 \\
SDSS J115820.75+043501.7 & $\cdots$ & sdL7\tablenotemark{g} & \cite{Kirkpatrick2014} &  2006 Apr 08 & 1080 & HD 97585 & 106 \\
\enddata
\tablenotetext{a}{Exact exposure times and which A0V stars were used could not be determined because the original FITS headers from the reduction were lost. Therefore we estimated the exposure times and standards using the raw data frames obtained form the IRTF Legacy Archive \href{http://irtfdata.ifa.hawaii.edu/search/}{http://irtfdata.ifa.hawaii.edu/search/}. }
\tablenotetext{b}{Calculated at the peak intensity in the $J$-band.}
\tablenotetext{c}{\cite{Kirkpatrick2010Stds} classify this object in the near-infrared as $<$esdM5.}
\tablenotetext{d}{\cite{Raiz2008} references \cite{RuizAnguita1993} for the spectral type, which provides a spectrum but no spectral type.}
\tablenotetext{e}{\cite{Burgasser2004} classify this object in the near-infrared as similar to, or slightly later than, sdM7.5.}
\tablenotetext{f}{\cite{Kirkpatrick2010Stds} classify this object in the near-infrared as L1 pec (blue).}
\tablenotetext{g}{\cite{Kirkpatrick2010Stds} classify this object in the near-infrared as sdL7.}
\end{deluxetable*}

\begin{figure*}
\centerline{\hbox{\includegraphics{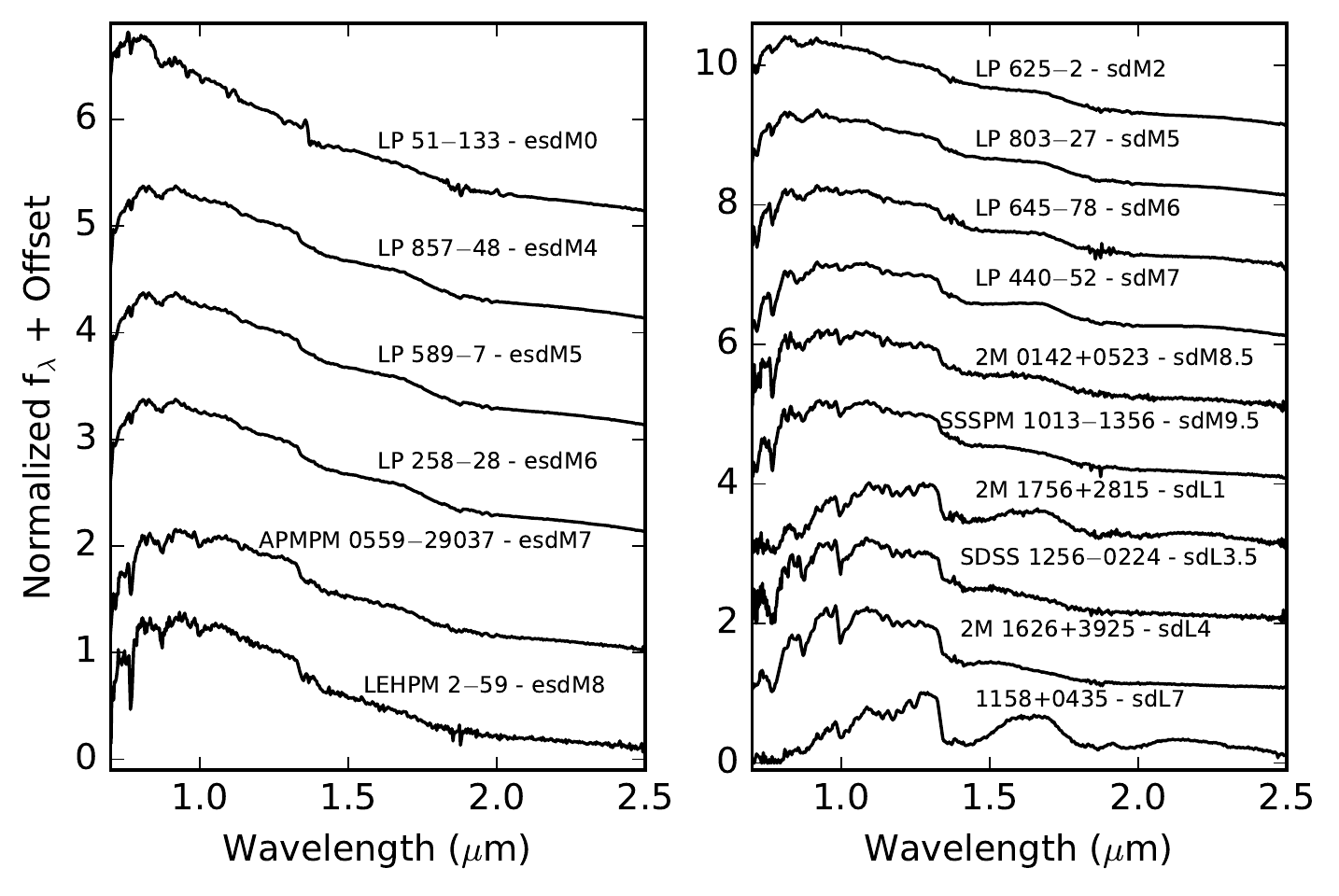}}}
\caption{Subdwarf standards, listed in Table 2.
\label{fig:sd_stds}}
\end{figure*}

\subsection{Magellan/FIRE}

Spectra of 5 objects were obtained with the Folded-Port Infrared Echellete (FIRE; \citealt{Simcoe2013FIRE}) spectrograph on the {\it Magellan} 6.5m Baade Telescope at Las Campanas Observatory. Observations were conducted on UT 18 Jul 2016. All observations were made with the high-throughput prism mode, which achieved a resolving power of R $\sim$ 450 across the 0.8--2.45 $\mu$m range. We used the $0\farcs6$-wide slit, aligned to the parallactic angle, and took exposures at two different nod positions along the slit. For all science targets, the sample-up-the-ramp mode was used. A0 V stars were observed after each science target to correct for telluric absorption and flux calibration. Data reduction was performed using a modified version of the Spextool reduction package (\citealt{Vacca2003Spex}; \citealt{Cushing2004Spex}). 

\subsection{CTIO/ARCoIRIS}
One object was observed on UT 09 Dec 2016 with Astronomy Research using the Cornell Infrared Imaging Spectrograph (ARCoIRIS) on the 4 m Blanco telescope located at the Cerro Tolo Inter-American Observatory (CTIO). ARCoIRIS takes simultaneous spectra across six cross-dispersed orders covering the 0.8--2.4$\mu$m range, with a resolving power of $\sim$3500. Science exposures were taken at two different nod positions along the slit, which has a fixed width of 1$''$. After observing our science target, we observed an A0 V star to use for telluric corrections and flux calibration. Data reduction was performed using a modified version of the Spextool reduction package (\citealt{Vacca2003Spex}; \citealt{Cushing2004Spex}). 

\section{Results}

\subsection{Spectral Classification}

Spectral types were determined by comparing each spectrum to the near-infrared spectral standards from \cite{Kirkpatrick2010Stds} and the near-infrared M and L subdwarf standards given in Table 2. First, the standard and object spectra were normalized to unity between 1.27 and 1.29 $\mu$m. One of us (J.J.G.) then assigned spectral types by eye, based on which spectral standard was the best match to each object over the 0.9--1.4 $\mu$m wavelength range. Some spectra fall appreciably red or blue of the spectral standards in the $H$- and $K$-bands and these are typed as `red' or `blue' respectively. Another of us (A.C.S.) confirmed all spectral types by eye, and the results are listed in Table 3.

 In total, we present spectra of 31 new M dwarfs, 18 new L dwarfs, and 11 new T dwarfs. Spectra of one additional L dwarf and four additional T dwarfs are also presented, but these have been previously published as discussed below. 13 of our objects are subdwarfs, including 9 new M subdwarfs and 4 new L subdwarfs. 11 of these objects (including two subdwarfs) with spectral types ranging from M7 to T7 are predicted to be within 25 pc. 


Five of our 65 objects have previously published spectra. \cite{Best2015} published a spectrum of WISE 0135+0205 (PSO J023.8557+02.0884), classifying it as an L9.5. We classified it as T0 (sl. red). \cite{Best2015} observed WISE 0316+2650 (PSO J049.1159+26.8409), classifying it as T2.5, with strong potential of being a binary. We classified it as a T3. \cite{Tinney2018} published a spectrum of WISE 1735$-$8209, classifying it as a T8; we classify it as a T7. \cite{LuhmanSheppard2014} observed WISE 2111$-$5211, classifying it as a T2.5; we typed it as a T3. \cite{Best2015} observed WISE 2249$-$1627 (PSO J342.3797$-$16.4665), classifying it as an L5, possibly in a binary with a T dwarf. \cite{SIMP2016} also observed this object, classifying it as an L4/T1 binary. We classified this as an L5 (blue). 

Additionally, two objects have published spectral types estimated using photometry. \cite{Tinney2018} used methane imaging to spectral type WISE0309$-$5016 as a T7, which agrees with our spectral type. \cite{Kirkpatrick2019} note that, because this object is much brighter in $M_H$ than other objects of similar $H-W2$ color, and much brighter in $M_{W1}$ and M$_\text{ch1}$ than other objects of similar {\it Spitzer} ch1$-$ch2 color, it is likely an unresolved binary. \cite{Kirkpatrick2019} estimated the spectral type of WISE0323$-$5907 based on {\it Spitzer} ch1 and ch2 photometry to be a T6. We classified this object as a T7. 

Finally we note that, in Table 10 of \cite{Schneider2016}, the spectral type of WISE 0413+2103 was mistaken for that of WISE 0413$-$2023. This caused WISE 0413+2103 to be listed as a late-type candidate, when it is in fact an M dwarf. We noticed this while selecting our follow-up candidates, and so observed WISE 0413$-$2023, which has a spectral type of L5 (blue).

\startlongtable
\begin{deluxetable*}{cccc}
\tablewidth{0pt}
\tablecaption{Spectral Types}
\label{table:properties}
\tablehead{
 \colhead{AllWISE Designation} \vspace{-0.25cm} & \colhead{Photometric\tablenotemark{a}} & \colhead{Spectral Type\tablenotemark{b}} & \colhead{Follow-Up\tablenotemark{c}}\\
 \colhead{} \vspace{-0.05cm} & \colhead{Spectral Type} & \colhead{From Observations} & \colhead{Category}}
\startdata
J000430.66$-$260402.3 & 20.5                   & T2 (blue)    & l      \\
J000458.47$-$133655.1 & 16.9                   & T2          & g      \\
J000536.63$-$263311.8 & 17.1                   & T0 (pec)     & l      \\
J000856.39$-$281321.7 & 18.0                   & L8          & n,l    \\
J010134.83+033616.0   & 7.0                    & M7          & s      \\
J010631.20$-$231415.1 & 18.2                   & L9          & l      \\
J011049.18$+$192000.1 & 9.1                    & M8          & w      \\
J013525.38+020518.2   & 17.7                   & T0 (sl. red)  & l      \\
J022721.93+235654.3   & 19.4                   & L9          & n,l    \\
J030119.39$-$231921.1 & 20.5                   & T1 (sl. blue) & n,l    \\
J030919.70$-$501614.2 & T7-T9\tablenotemark{d} & T7          & n,l    \\
J031627.79+265027.5   & 19.0                   & T3          & l      \\
J032309.12$-$590751.0 & 26.2                   & T7          & n,l,s  \\
J032838.73+015517.7   & 18.5                   & L5 (blue)    & l      \\
J033346.88+385152.6   & 11.6                   & M8          & s      \\
J034409.71+013641.5   & 19.1                   & L8          & l      \\
J034858.75$-$562017.8 & 22.6                   & T3          & n,l    \\
J041353.96$-$202320.3 & 17.7                   & L5 (blue)    & g      \\
J041743.13+241506.3   & 23.7                   & T6          & n,l    \\
J053424.45+165255.0   & 15.4                   & L2 (pec)     & n      \\
J054455.54+063940.3   & 10.3                   & M9          & n      \\
J061429.77+383337.5   & 10.3                   & M9          & n      \\
J062858.69+345249.2   & 17.2                   & L4          & l      \\
J063552.52+514820.4   & 17.4                   & T0          & l      \\
J084254.56$-$061023.7 & 22.7                   & T4          & l,n    \\
J085039.11$-$022154.3 & 16.3                   & sdL7 (red)  & n      \\
J085633.87$-$181546.6 & 11.6                   & L1          & g      \\
J092453.76+072306.0   & 5.9                    & M6          & s      \\
J094812.21$-$290329.5 & 11.8                   & sdL1        & s      \\
J095230.79$-$282842.2 & 5.3                    & esdM4       & s      \\
J101944.62$-$391151.6 & 24.0                   & T3 (blue)    & n,l,s  \\
J103534.63$-$071148.2 & 17.7                   & sdL7        & l      \\ 
J111320.39+501010.5   & 7.0                    & M4 (blue)    & s      \\
J112158.76+004412.3   & 8.6                    & M7 (blue)    & s      \\
J112859.45+511016.8   & 14.1                   & L3          & g      \\
J120751.17+302808.9   & 10.6                   & M8          & s      \\
J121231.97$-$050750.7 & 5.4                    & M7 (sl. blue) & s      \\
J121914.75+081027.0   & $<$5                   & sdM7        & s      \\
J122042.20+620528.3   & 6.3                    & sdM7        & s      \\
J123513.87$-$045146.5 & 5.1                    & esdM4       & s      \\
J124516.66+601607.5   & 9.4                    & sdM8.5      & s      \\
J133520.09$-$070849.3 & 12.3                   & M7          & s      \\
J134359.71+634213.1   & 10.3                   & M8          & g      \\
J143942.79$-$110045.4 & 11.8                   & sdL1        & s      \\
J144056.64$-$222517.8 & 9.1                    & sdM8.5      & s      \\
J145645.54$-$103343.5 & 10.2                   & M8 (sl. blue) & g      \\
J145747.55$-$094719.3 & 6.5                    & esdM4       & s      \\
J155225.22+095155.5   & 7.9                    & sdM7        & s      \\
J165057.66$-$221616.8 & 5.4                    & M5          & n      \\
J171059.52$-$180108.7 & 5.2                    & M4(blue)    & n      \\
J171105.08$-$275531.7 & 7.3                    & M6          & n      \\
J171454.88+064349.8   & 15.0                   & L2(red)     & g      \\
J173551.56$-$820900.3 & 24.3                   & T7          & n,l    \\
J180839.55+070021.7   & 14.7                   & L1 (blue)    & s      \\
J182010.20+202125.8   & 7.4                    & sdM8.5      & s      \\
J183654.10$-$135926.2 & 8.7                    & M6          & n      \\
J191011.03+563429.3   & 11.6                   & M8          & n      \\
J201252.78+124633.3   & 6.5                    & M7 (sl. red)  & n      \\
J211157.84$-$521111.3 & 19.7                   & T3          & l      \\
J215550.34$-$195428.4 & 16.7                   & L7          & g      \\
J221737.41$-$355242.7 & 10.3                   & M5          & s      \\
J223444.44$-$230916.1 & 17.4                   & L5          & l      \\ 
J224931.10$-$162759.6 & 17.1                   & L5 (blue)    & l      \\
J230743.63+052037.3   & 11.7                   & M7          & w      \\
J234404.85$-$250042.2 & 11.8                   & M7          & s      \\
\enddata
\tablenotetext{a}{Estimated spectral types are from \cite{Schneider2016}. They are numerical spectral types where, for example, M2=2, L2=12, T2=25, etc.}
\tablenotetext{b}{Spectral types as determined by comparing our SpeX Prism spectra with spectral standards. Subdwarf spectral types are denoted by the following abbreviations: sd = subdwarf, d/sd = dwarf/subdwarf, esd = extreme subdwarf.}
\tablenotetext{c}{Lists which of our follow-up categories an object belonged to: n=nearby, s=subdwarf, l=late-type, g=gap object, w=poor weather target.}
\tablenotetext{d}{WISE 0309$-$5016 does not show up in 2MASS, so a precise numerical spectral type could not be determined. Instead, an estimate was made based on the W1-W2 color. See Section 3.4 of \cite{Schneider2016} for details.}
\end{deluxetable*}

\subsection{Distance Estimates}

We can improve upon the spectrophotometric distances of \cite{Schneider2016} by using available photometry and absolute magnitude-spectral type relations to compute spectroscopic distances for each of our objects. We primarily used the relations of \cite{DupuyLiu2012}, which are valid for objects with spectral types between M6 and T9 (inclusive) and can be used with 2MASS $J$, $H$, and $K_s$ and WISE $W1$ and $W2$ photometry. For spectral types earlier than M6, we used the relations of \cite{Zhang2013}, which are valid for spectral types between M1 and L9 (inclusive) and can be used with 2MASS $J$, $H$, and $K_s$ photometry. Finally, for the subdwarfs, we used the relations of \cite{Zhang2017}, which are valid for subdwarfs with spectral types between M0 and L7 (inclusive) and can be used with 2MASS $J$ and $H$ band photometry. 

These relations were combined with available photometry to calculate the distances and their uncertainties using a Monte Carlo approach to properly account for the uncertainties in the spectral type, spectral type-absolute magnitude relation, and the photometry. We randomly drew from distributions for the spectral type, the absolute magnitude, and the apparent magnitude to compute a distance.  A uniform distribution with a width of 1 subtype centered on the spectral type of the object was used for the spectral type distribution, a normal distribution with a mean and standard deviation given by the spectral-type absolute magnitude relation and RMS uncertainty of that relation was used for the absolute magnitude relation, and a normal distribution with a mean and standard deviation given by the apparent magnitude and its uncertainty was used for the apparent magnitude distribution.  The process was repeated 10,000 times for each object, and the mean and standard deviation of the resulting distribution gave us the spectroscopic distance and its uncertainty. Distances and uncertainties were calculated for each object in the filters where the spectral type - absolute magnitude relations are valid, and a weighted average of all individual spectroscopic distances for each object was then used to calculate the final spectroscopic distances, which can be found in Table 4. 

\startlongtable
\begin{deluxetable*}{lllllll}
\tablewidth{0pt}
\tablecaption{Object Distances}
\tabletypesize{\scriptsize}
\label{table:properties}
\tablehead{
\colhead{AllWISE Designation} \vspace{-0.35cm} & \colhead{Sp\tablenotemark{a}} & \colhead{Schneider 2016\tablenotemark{b}} & \colhead{Our}        & \colhead{Gaia Source\tablenotemark{c}} & \colhead{Gaia\tablenotemark{d}} & \colhead{Kirkpatrick 2018\tablenotemark{e}} \\
         \colhead{}  \vspace{-0.005cm}         & \colhead{Type}                & \colhead{Dist (pc)}                       & \colhead{Dist (pc)}   & \colhead{ID}                           & \colhead{Dist (pc)}             & \colhead{Dist (pc)}}
\startdata
J000430.66$-$260402.3 & T2 (blue)    & $\cdots$ & 25  $\pm$ 2.3  & $\cdots$            & $\cdots$                     & $\cdots$         \\
J000458.47$-$133655.1 & T2          & $\cdots$ & 29  $\pm$ 2.7  & $\cdots$            & $\cdots$                     & $\cdots$         \\
J000536.63$-$263311.8 & T0 (pec)     & $\cdots$ & 31  $\pm$ 2.6  & $\cdots$            & $\cdots$                     & $\cdots$         \\
J000856.39$-$281321.7 & L8          & 24--34   & 29  $\pm$ 2.5  & $\cdots$            & $\cdots$                     & $\cdots$         \\
J010134.83+033616.0   & M7          & $\cdots$ & 92  $\pm$ 7.7  & 2551477793805008256 & $ 86  _{ -5.1}^{ +5.8}$ & $\cdots$         \\
J010631.20$-$231415.1 & L9          & $\cdots$ & 36  $\pm$ 3.2  & $\cdots$            & $\cdots$                     & $\cdots$         \\
J011049.18$+$192000.1 & M8          & $\cdots$ & 59  $\pm$ 4.9  & 2786913366801779968 & $ 51.5  _{ -0.95}^{ +0.98}$ & $\cdots$         \\
J013525.38+020518.2   & T0 (sl. red)  & $\cdots$ & 25  $\pm$ 2.1  & $\cdots$            & $\cdots$                     & $\cdots$         \\
J022721.93+235654.3   & L9          & 22--31   & 28  $\pm$ 2.3  & $\cdots$            & $\cdots$                     & $\cdots$         \\
J030119.39$-$231921.1 & T1 (sl. blue) & 24--33   & 27  $\pm$ 2.3  & $\cdots$            & $\cdots$                     & $\cdots$         \\
J030919.70$-$501614.2 & T7          & 9--13    & 14  $\pm$ 1.8  & $\cdots$            & $\cdots$                     & 15.0 $\pm$ 0.87 \\
J031627.79+265027.5   & T3          & $\cdots$ & 22  $\pm$ 2.1  & $\cdots$            & $\cdots$                     & $\cdots$         \\
J032309.12$-$590751.0 & T7          & 16--26   & 19  $\pm$ 2.2  & $\cdots$            & $\cdots$                     & 14.0 $\pm$ 0.84 \\
J032838.73+015517.7   & L5 (blue)    & $\cdots$ & 47  $\pm$ 4.5  & $\cdots$            & $\cdots$                     & $\cdots$         \\
J033346.88+385152.6   & M8          & $\cdots$ & 104 $\pm$ 8.7  & 236441149397820800  & $ 85  _{ -7.2}^{ +8.6}$ & $\cdots$         \\
J034409.71+013641.5   & L8          & $\cdots$ & 37  $\pm$ 3.5  & $\cdots$            & $\cdots$                     & $\cdots$         \\
J034858.75$-$562017.8 & T3          & 24--33   & 28  $\pm$ 3.0  & $\cdots$            & $\cdots$                     & $\cdots$         \\
J041353.96$-$202320.3 & L5 (blue)    & $\cdots$ & 41  $\pm$ 3.5  & $\cdots$            & $\cdots$                     & $\cdots$         \\
J041743.13+241506.3   & T6          & 13--19   & 12  $\pm$ 1.0  & $\cdots$            & $\cdots$                     & $\cdots$         \\
J053424.45+165255.0   & L2 (pec)     & 18--25   & 33  $\pm$ 2.8  & 3397015189186833408 & $ 28  _{ -2.8}^{ +3.6}$ & $\cdots$         \\
J054455.54+063940.3   & M9          & 24--37   & 35  $\pm$ 2.9  & 3333278694852547328 & $ 31.3  _{ -0.35}^{ +0.36}$ & $\cdots$         \\
J061429.77+383337.5   & M9          & 18--27   & 27  $\pm$ 2.2  & 956200977271782144  & $ 25.2  _{ -0.23}^{ +0.24}$ & $\cdots$         \\
J062858.69+345249.2   & L4          & $\cdots$ & 40.  $\pm$ 3.3  & $\cdots$            & $\cdots$                     & $\cdots$         \\
J063552.52+514820.4   & T0          & $\cdots$ & 29  $\pm$ 2.7  & $\cdots$            & $\cdots$                     & $\cdots$         \\
J084254.56$-$061023.7 & T4          & 20--29   & 21  $\pm$ 1.9  & $\cdots$            & $\cdots$                     & $\cdots$         \\
J085039.11$-$022154.3 & sdL7 (red)  & 21--30   & 24  $\pm$ 3.4  & $\cdots$            & $\cdots$                     & $\cdots$         \\
J085633.87$-$181546.6 & L1          & $\cdots$ & 64  $\pm$ 5.4  & 5728941156831133952 & $ 56  _{ -3.5}^{ +4.0}$ & $\cdots$         \\
J092453.76+072306.0   & M6          & $\cdots$ & 140 $\pm$ 12 & 586424457955450496  & $118  _{ -7.3}^{ +8.3}$ & $\cdots$         \\
J094812.21$-$290329.5 & sdL1        & $\cdots$ & 71  $\pm$ 9.8  & 5656672112963964928 & $ 62  _{ -2.6}^{ +2.8}$ & $\cdots$         \\
J095230.79$-$282842.2 & esdM4       & $\cdots$ & 110 $\pm$ 15 & 5464936251656505344 & $130.  _{ -2.5}^{ +2.6}$ & $\cdots$         \\
J101944.62$-$391151.6 & T3 (blue)    & 19--28   & 23  $\pm$ 2.0  & $\cdots$            & $\cdots$                     & $\cdots$         \\
J103534.63$-$071148.2 & sdL7        & $\cdots$ & 42  $\pm$ 6.1  & $\cdots$            & $\cdots$                     & $\cdots$         \\ 
J111320.39+501010.5   & M4 (blue)    & $\cdots$ & 140 $\pm$ 19 \tablenotemark{f} & 838162769031557888  & $181  _{ -9.3}^{+10.4}$ & $\cdots$         \\
J112158.76+004412.3   & M7 (blue)    & $\cdots$ & 116 $\pm$ 9.9  & 3798149260432886528 & $ 75  _{ -6.7}^{ +8.2}$ & $\cdots$         \\
J112859.45+511016.8   & L3          & $\cdots$ & 46  $\pm$ 3.8  & $\cdots$            & $\cdots$                     & $\cdots$         \\
J120751.17+302808.9   & M8          & $\cdots$ & 78  $\pm$ 6.5  & 4014105473115624192 & $ 71  _{ -3.5}^{ +3.9}$ & $\cdots$         \\
J121231.97$-$050750.7 & M7 (sl. blue) & $\cdots$ & 71  $\pm$ 6.0  & 3596616230830390016 & $ 66  _{ -1.8}^{ +1.9}$ & $\cdots$         \\
J121914.75+081027.0   & sdM7        & $\cdots$ & 120 $\pm$ 16 & 3902112585964749312 & $122  _{ -7.5}^{ +8.5}$ & $\cdots$         \\
J122042.20+620528.3   & sdM7        & $\cdots$ & 88  $\pm$ 7.4  & 1583395326382043392 & $113  _{ -4.0}^{ +4.3}$ & $\cdots$         \\
J123513.87$-$045146.5 & esdM4       & $\cdots$ & 160 $\pm$ 22 & 3680363115235579904 & $156  _{ -4.9}^{ +5.2}$ & $\cdots$         \\
J124516.66+601607.5   & sdM8.5      & $\cdots$ & 100 $\pm$ 14 & 1579775596664490752 & $116  _{ -3.7}^{ +4.0}$ & $\cdots$         \\
J133520.09$-$070849.3 & M7          & $\cdots$ & 130 $\pm$ 11 & 3630793763800277376 & $ 100.  _{ -9.4}^{+11}$ & $\cdots$         \\
J134359.71+634213.1   & M8          & $\cdots$ & 98  $\pm$ 8.3  & 1665037775596252544 & $ 80.  _{ -5.8}^{ +6.8}$ & $\cdots$         \\
J143942.79$-$110045.4 & sdL1        & $\cdots$ & 80.  $\pm$ 11 & 6324908688520221568 & $130  _{-26}^{+47}$ & $\cdots$         \\
J144056.64$-$222517.8 & sdM8.5      & $\cdots$ & 80.  $\pm$ 11 & 6278872445902622336 & $106  _{ -3.8}^{ +4.0}$ & $\cdots$         \\
J145645.54$-$103343.5 & M8 (sl. blue) & $\cdots$ & 61  $\pm$ 5.0  & 6313890619936907136 & $ 49  _{ -1.3}^{ +1.4}$ & $\cdots$         \\
J145747.55$-$094719.3 & esdM4       & $\cdots$ & 140 $\pm$ 19 & 6326026685686833920 & $166  _{ -5.5}^{ +5.9}$ & $\cdots$         \\
J155225.22+095155.5   & sdM7        & $\cdots$ & 130 $\pm$ 18 & 4455454422667645184 & $130  _{-12}^{+14}$ & $\cdots$         \\
J165057.66$-$221616.8 & M5          & 22--35   & 41  $\pm$ 6.4  & 4126600390415016832 & $ 34.7  _{ -0.12}^{ +0.12}$ & $\cdots$         \\
J171059.52$-$180108.7 & M4 (blue)    & 24--37   & 60  $\pm$ 12 & 4134686886136743552 & $ 44.2  _{ -0.17}^{ +0.18}$ & $\cdots$         \\
J171105.08$-$275531.7 & M6          & 21--34   & 33  $\pm$ 2.8  & $\cdots$            & $\cdots$                     & $\cdots$         \\
J171454.88+064349.8   & L2 (red)     & $\cdots$ & 56  $\pm$ 4.8  & $\cdots$            & $\cdots$                     & $\cdots$         \\
J173551.56$-$820900.3 & T7          & 14--21   & 13  $\pm$ 1.4  & $\cdots$            & $\cdots$                     & 13.3 $\pm$ 0.81 \\
J180839.55+070021.7   & L1 (blue)    & $\cdots$ & 79  $\pm$ 6.7  & $\cdots$            & $\cdots$                     & $\cdots$         \\
J182010.20+202125.8   & sdM8.5      & $\cdots$ & 80.  $\pm$ 11 & 4528661276939071488 & $124  _{ -3.0}^{ +3.1}$ & $\cdots$         \\
J183654.10$-$135926.2 & M6          & 20--31   & 35  $\pm$ 3.0  & $\cdots$            & $\cdots$                     & $\cdots$         \\
J191011.03+563429.3   & M8          & 16--23   & 29  $\pm$ 2.4  & 2141364423410899968 & $ 23.48  _{ -0.07}^{ +0.07}$ & $\cdots$         \\
J201252.78+124633.3   & M7 (sl. red)  & 17--26   & 20.  $\pm$ 1.7  & 1803225427774999680 & $ 19.27  _{ -0.03}^{ +0.03}$ & $\cdots$         \\
J211157.84$-$521111.3 & T3          & $\cdots$ & 26  $\pm$ 2.4  & $\cdots$            & $\cdots$                     & $\cdots$         \\
J215550.34$-$195428.4 & L7          & $\cdots$ & 39  $\pm$ 3.6  & $\cdots$            & $\cdots$                     & $\cdots$         \\
J221737.41$-$355242.7 & M5          & $\cdots$ & 150 $\pm$ 24 & $\cdots$            & $\cdots$                     & $\cdots$         \\
J223444.44$-$230916.1 & L5          & $\cdots$ & 55  $\pm$ 5.2  & $\cdots$            & $\cdots$                     & $\cdots$         \\ 
J224931.10$-$162759.6 & L5 (blue)    & $\cdots$ & 37  $\pm$ 3.1  & $\cdots$            & $\cdots$                     & $\cdots$         \\
J230743.63+052037.3   & M7          & $\cdots$ & 64  $\pm$ 5.4  & 2662702873947256832 & $ 87  _{ -2.7}^{ +2.9}$ & $\cdots$         \\
J234404.85$-$250042.2 & M7          & $\cdots$ & 84  $\pm$ 7.2  & 2338610933917661696 & $ 63  _{ -5.1}^{ +6.0}$ & $\cdots$         \\
\enddata
\tablenotetext{a}{Spectral types as determined by comparing our SpeX Prism spectra with spectral standards. Subdwarf spectral types are denoted by the following abbreviations: sd = subdwarf, d/sd = dwarf/subdwarf, esd = extreme subdwarf.}
\tablenotetext{b}{Only listed for objects that were nearby candidates in the \cite{Schneider2016} survey}
\tablenotetext{c}{Only listed for objects with matches in Gaia DR2}
\tablenotetext{d}{Taken from \cite{Bailer-Jones2018}.}
\tablenotetext{e}{Only listed for three objects, which were included in the \cite{Kirkpatrick2019} paper.}
\tablenotetext{f}{This is the distance we calculated, assuming this object is a subdwarf. In the absence of this assumption, the distance would be 260 $\pm$ 31 pc.}
\end{deluxetable*}


Most of our distances are within, or close to, the distance ranges from the \cite{Schneider2016} survey. WISE 1710$-$1801 shows a large discrepancy between the spectroscopic distance calculated in this paper (60 $\pm$ 12 pc) and the spectrophotometric distance estimated in \cite{Schneider2016} (24--37 pc). This is likely a result of the fact that \cite{Schneider2016} used the \cite{DupuyLiu2012} relations to calculate their distance estimate, and these relations are not valid for early M dwarfs. The estimate in this paper used the \cite{Zhang2013} relations, which are valid for early M dwarfs. 


We also searched the Gaia DR2 archive to identify which of our candidates were detected by Gaia. Using the 2MASS - AllWISE proper motions calculated by \cite{Schneider2016}, and the positions of our sources from the AllWISE epoch (2010.5), we calculated the positions of each of our sources in the Gaia epoch (2015.5).  It was not possible to do this for WISE 0309$-$5016, because it was not detected in 2MASS and \cite{Schneider2016} were not able to calculate a 2MASS - AllWISE proper motion for it. We then cross-matched the positions of our objects at the Gaia epoch against the Gaia DR2 archive, and identified all Gaia matches within 5$\arcsec$. We then examined all the matches for each object to confirm matches, and in some cases, determine which of multiple matches was the correct object. This was accomplished by: first, performing a visual inspection of each of our objects using finder charts, examining the position of our object in images from DSS, UKIDSS, 2MASS, WISE, and Pan-STARRs, where available. Second, the separation was calculated between the coordinates we calculated for each object at the Gaia epoch and the coordinates for each match in the Gaia DR2 catalog, to determine which of the multiple matches was closest to the coordinates we calculated. Third, we compared the proper motions for each match in the Gaia Catalog to the proper motions for each source calculated in \cite{Schneider2016}, to make sure those values matched. 

\begin{figure}
\centerline{\hbox{\includegraphics[width=\columnwidth]{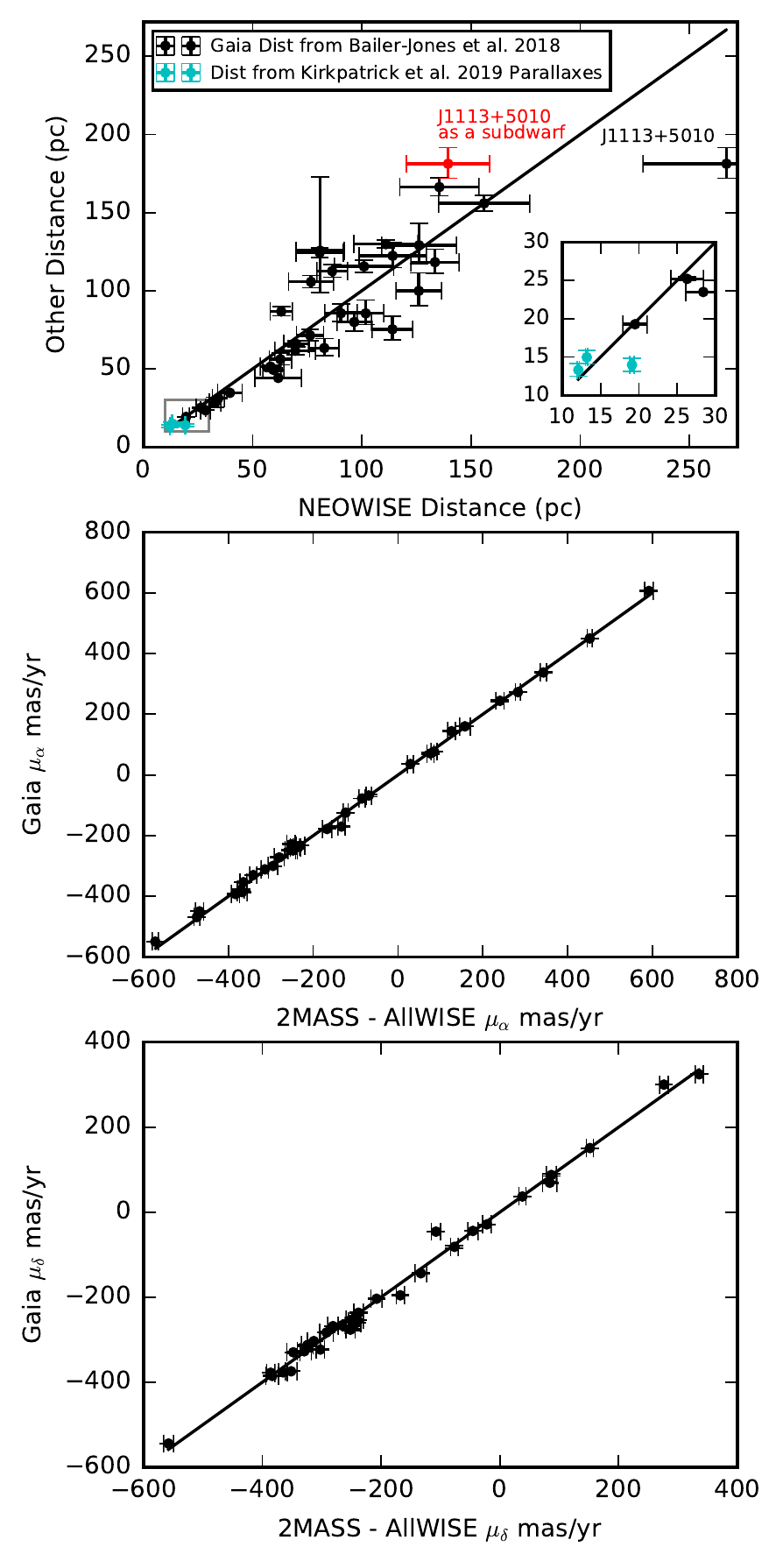}}}
\caption{Top panel: Comparison of the spectroscopic distances we calculated for each of our objects to the distances from Gaia, as determined by \cite{Bailer-Jones2018} and the distances determined from the parallaxes of \cite{Kirkpatrick2019}. One of our objects, WISE 1113+5010 shows a large discrepancy between the NEOWISE distance and the Gaia distance. We believe this is because it may be a subdwarf (see \S4.2 for details). We estimated what the spectroscopic distance would be if it was a subdwarf, and that matches up much better with the Gaia distance, as show in this figure. Middle and bottom panels: comparison of the NEOWISE proper motions and the Gaia proper motions for all objects that appear in Gaia. There is good agreement between these for all of our objects. \label{fig:distcomp}}
\end{figure}

In total, 32 of our 65 objects have matches in Gaia. They are all listed in Table 4, along with the Gaia source ID for each match, and the Gaia distances for each object \citep{Bailer-Jones2018}. Two of our objects (WISE 0850$-$0221 and WISE 1808+0700) had matches in the Gaia catalog with no parallax measurements, and so are not included in Table 4. For most of our objects, we find good agreement between our spectroscopic distances and the Gaia distances, as well as the 2MASS - AllWISE proper motions and the Gaia proper motions, as can be seen in Figure \ref{fig:distcomp}. For WISE 1113+5010, we noticed a large discrepancy between our spectroscopic distance of 260 $\pm$ 31 pc and the Gaia distance of  $181  _{ -9.3}^{+10.4}$ pc. Our spectral type for this object is an M4 (blue), meaning it exhibits suppressed flux in the $H$- and $K$-bands, relative to the $J$-band, causing it to appear bluer in the $H$- and $K$- bands than field objects of the same spectral class. This is typically an indicator that an object could be a subdwarf (this is discussed in greater detail in \S 5.3). If we use the absolute magnitude-spectral type relations for subdwarfs, we get a distance of 140$\pm$19 pc, which is much closer to the Gaia distance. This suggests that WISE 1113+5010 may either be a subdwarf (sdM4) or an intermediate subdwarf (d/sdM4). Unfortunately, we do not have a spectrum of a sdM4 standard, and so we cannot confirm this hypothesis.

Included in Table 4 along with our Gaia distances, are distances for three objects (WISE 0309$-$5016, WISE 0323$-$5907 and WISE 1735$-$8209) calculated from parallaxes obtained by \cite{Kirkpatrick2019} using the The Infrared Array Camera (IRAC; \citealt{Fazio2004}) on the {\it Spitzer Space Telescope} \citep{Werner2004Spitzer}. The distances for WISE 0309$-$5016 and WISE 1735$-$8209 agree very well with the spectroscopic distances we calculated, but the distance for WISE 0323$-$5907 does not, as can be seen in Figure \ref{fig:distcomp}. The parallax from \cite{Kirkpatrick2019} gives a distance of 14.0 $\pm$ 0.84 pc, and our spectroscopic distance is 19 $\pm$ 2.1 pc. The reason for the large discrepancy is still unclear. \citeauthor{Kirkpatrick2019} estimate the spectral type of this object to be T6, based on the ch1$-$ch2 photometry but, according to our spectrum from IRTF/SpeX, it is a textbook T7. \citeauthor{Kirkpatrick2019} note that this source is too faint in $W1$ and $W2$ for its Spitzer ch1$-$ch2 color. If we calculate the distance using only the 2MASS $J$-band photometry (which comes from the 2MASS reject catalog), we get a distance of 18.1 $\pm$ 4.2 pc, which falls within 1$\sigma$ of the \citeauthor{Kirkpatrick2019} value. This object will need to be studied further to determine the exact reason for this discrepancy.

\section{Discussion}

\subsection{Nearby Objects}






Volume-limited samples are the gold standard in astrophysics because they provide an unbiased sample of the objects under scrutiny. Constructing a complete census of the stars and brown dwarfs in the solar neighborhood is particularly important because this region contains the brightest, and thus most easily studied, objects of a given spectral class. At least one star or brown dwarf has been added to the list of stellar systems that lie within 10 pc of the Sun every year since 2002 \citep{Henry2018} indicating that the local census remains incomplete. The intrinsic faintness of brown dwarfs makes constructing volume-limited samples difficult, particularly out to larger distances where the census is even more incomplete.

Our survey and follow-up observations have identified 21 new objects within 30 pc of the Sun. Eleven of these objects have distances within 25 pc: one M dwarf (WISE 2012+1246 M7  (sl. red)), nine T dwarfs (WISE 0004$-$2604 T2 (blue),  WISE 0135+0205 T0  (sl. red), WISE 0309$-$5016 T7, WISE 0316+2650 T3, WISE 0323$-$5907 T7, WISE 0417+2415 T6, WISE 0842$-$0610 T4, WISE 1019$-$3911 T3 (blue), and WISE 1735$-$8209 T7), and one L subdwarf (WISE 0850$-$0221 sdL7 (red)). An additional ten objects have spectroscopic distances 25 pc $<$ d $<$ 30 pc: two M dwarfs (WISE 0614+3833 M9, WISE 1910+5634 M8), two L dwarfs (WISE 0008$-$2813 L8, WISE 0227+2356 L9), and six T dwarfs (WISE 0004$-$1336 T2, WISE 0301$-$2319 T0 (sl. blue), WISE 0328$-$5620 T3, WISE 0348$-$5620 T3, WISE 0635+5148 T0, WISE 2111-5211 T3). Three of our objects are within 15 pc. All three of these are T dwarfs: (WISE 0309$-$5016 (13.8 $\pm$1.69; T7), WISE 1735$-$8209 (12.4$\pm$1.28; T7), WISE 0417+2415 (11.4$\pm$0.96; T6)). Even though it is within 15pc, WISE 0417+2415 has no published parallax. The other two have parallaxes published in \cite{Kirkpatrick2019}.

\subsection{Late-type Brown Dwarfs}




While the stellar mass function in the solar neighborhood is well understood \citep{Bastian2010}, the substellar mass function has proven more difficult to measure for two reasons. First, brown dwarfs cool over time, and thus do not follow a mass-luminosity relation as stars do. Second, as mentioned in \S 5.1, the census of brown dwarfs in the solar neighborhood remains incomplete. The census is most incomplete for the late-type T and Y dwarfs because of their intrinsic faintness. However, these objects are among the most important because it has been shown that they provide the best constraints on the underlying mass function (e.g. \citealt{Burgasser2004IMF} and \citealt{Kirkpatrick2019}).

In an effort to identify new late-type objects in the solar neighborhood, we observed 23 candidate late-type objects ($\geq$L7) from \cite{Schneider2016}. Fourteen of these are T dwarfs, with spectral types ranging from T0 to T7; four are late-type L dwarfs; and one (WISE 1035$-$0711) is a sdL7. The remaining four were mid-L dwarfs, with spectral types of either L4 or L5. We also discovered three additional late-type objects: WISE 2155$-$1954 (L7), WISE 0004$-$1336 (T2), and WISE 0850$-$0221 (sdL7 (red)), which were not late-type candidates.

WISE 0004$-$1336 was one of the objects we observed to fill in gaps in our Right Ascension coverage (see \S2). In \cite{Schneider2016} it had an estimated spectral type, based on the available photometry, of L6.9, making it just beyond the L7 cutoff, so it was not listed as one of their late-type candidates. When we observed this object, we found it to have a spectral type of T2. Three other objects (WISE 0005$-$2633; WISE 0135+0205; WISE 0635+5148) also had estimated spectral types, based on photometry, of L7 or earlier, and were classified as T dwarfs based on their spectra. In \cite{Schneider2016}, the spectral types for these objects were estimated based on their infrared colors (see Appendix of \cite{Schneider2016} for details), using available photometry. The colors for early T dwarfs can overlap with the colors of mid- to early-L dwarfs (see Figure 5 in \cite{Schneider2016}), which can cause these objects to be mistakenly classified as mid-L dwarfs because of the similarity in color. It is likely that this is why these objects were mis-classified as L6 or L7, instead of T dwarfs, and why WISE 0004$-$2604 missed the cut-off for the late-type objects in \cite{Schneider2016}.


\subsection{Subdwarfs}

While effective temperature is the primary parameter that controls the spectral morphology of brown dwarfs, both surface gravity and metallicity also play a role. Our understanding of the impacts that variations in metallicity have on the emergent spectra of brown dwarfs is still in its infancy because of the paucity of metal-poor L and T dwarfs known; the total number currently stands at 71 \citep{Zhang2018} which in in stark contrast to the thousands of near solar metallicity brown dwarfs known. Identifying new metal-poor brown dwarfs will help us to build a large enough sample to begin inferring trends in spectral morphology within a given spectral type, and will allow us to better examine trends across a larger range of subdwarf types.


We conducted follow-up observations of 24 candidate subdwarfs from \cite{Schneider2016}. As described in \S 3.1, we have spectra of 16 M and L subdwarf standards, obtained with IRTF/SpeX, which we used to determine which of our objects were subdwarfs. While this is the first time these have been used as near-infrared subdwarf standards, all had previously been spectral typed as subdwarfs in the optical, as detailed in Table 2. Our subdwarf spectral standards include both sd and esd for the M spectral class, and sd for the L spectral class. The esd have $-$1.7 $<$ [Fe/H] $\leq$ $-$1.0, and the sd have $-$1.0 $<$ [Fe/H] $\leq$ $-$0.3 (\citealt{Gizis1997}; \citealt{Zhang2017}). All of our observed objects, both subdwarf candidates and non-subdwarf candidates were compared against both the subdwarf and non-subdwarf standards during the spectral typing process. Final spectral types were determined based on the best match between each object and all available spectral standards. As can be see in Table 2 and Figure 7, our spectral sequence of subdwarf standards is incomplete, especially for the L subdwarfs. This is due to the fact that, at present, there are very few near-infrared spectral standards for subdwarfs available. We have spectral typed our objects to the best of our ability with the available standards, but, we have likely missed some of the subdwarfs in our sample, as a result of not having standards at those spectral types. 

Of the 24 subdwarf candidates we observed, 11 were spectral typed as subdwarfs: six sdM (WISE 1219+018 sdM7; WISE 1220+6205 sdM7; WISE 1245+6016 sdM8.5; WISE 1440$-$2225 sdM8.5; WISE 1552+0951 sdM7; WISE 1820+2021 sdM8.5), two sdL (WISE 0948$-$2903 sdL1; WISE 1439$-$1100 sdL1), and three esdM4 (WISE 0952$-$2828; WISE 1235$-$0451; WISE 1457$-$0947). Ten of the remaining 13 were spectral typed as M dwarfs, with spectral types ranging from M4 to M8, one is an L1 (blue) (WISE 1808+0700), and one is a T7 (WISE 0323$-$5907). The remaining object, WISE 1019$-$3911, was spectral typed as a T3 (blue). We also observed two objects that were not subdwarf candidates, but were spectral typed as subdwarfs. Both are L subdwarfs. WISE 0850$-$0221 is a sdL7 (red) and WISE 1035$-$0711 is a sdL7. According to \cite{Zhang2018}, there are 66 known L subdwarfs. This includes four sdL7s, three sdL5s and four sdL1. We have discovered two additional sdL7s, two additional sdL1s and three candidate sdL5s, substantially increasing the number of known L subdwarfs at these spectral types.

Due to enhanced collision-induced $H_2$ absorption, subdwarfs tend to have suppressed flux in the $H$- and $K$-bands, relative to the $J$-band, causing them to appear bluer in the $H$- and $K$- bands than field objects of the same spectral class. In addition, they exhibit brightening in the $Y$-band. Among the objects we observed, 11 are blue: four M dwarfs, four L dwarfs, and three T dwarfs. For three of these (WISE 1121+0044, M7 (blue); WISE 1456$-$1033 M8 (sl. blue), and WISE 1808+0700 L1 (blue)), we have subdwarf spectral standards at those spectral types and so can confirm that, while they are blue, they are not subdwarfs. For the remaining eight, we do not. We believe these objects could be subdwarf candidates, but without subdwarf standards at the corresponding spectral types, we cannot be certain at this time. We have three new candidate T subdwarfs: WISE 0301$-$2319 (sdT1), WISE 0004$-$2604 (sdT2), and WISE 1019$-$3911 (sdT3); three new candidate sdL5: WISE 0328+0155, WISE 0413$-$2023, and WISE 2249$-$1627; and two new candidate sdM4: WISE 1113+5010 and WISE 1710$-$1801.

Additionally, WISE 0948$-$2903 (sdL1), and WISE 1439$-$1100 (sdL1), and some of the blue late Ms (e.g., WISE 1212$-$0507 and WISE 1121+0044) show a triangular $H$-band peak, a feature that is seen in the spectra of young, low-gravity M and L dwarfs and attributed to reduced collision induced $H_2$ absorption in low pressure atmospheres (e.g. \citealt{Rice2011}; \citealt{AllersLiu2013}). \cite{Aganze2016} analyzed this feature while studying the d/sdM7 GJ 660.1B which has [Fe/H] = $-$0.63$\pm$0.06, and found that this feature is also indicative of subsolar metallicity. The presence of this feature in our spectra supports the classification of WISE 0948$-$2903 and WISE 1439$-$1100 as subdwarfs, and suggests the blue M dwarfs may also have subsolar metallicities.

Among our discoveries, we find three new T subdwarf candidates all with distances around 25pc. One of these, WISE 1019$-$3911, was listed as a candidate in all three categories. Based on the estimates from \cite{Schneider2016}, WISE 1019$-$3911 was expected to be a T dwarf, with an estimated spectral type based on photometry of T4, an estimated distance of 19--28 pc, and was also a subdwarf candidate. We observed it using CTIO/ARCoIRIS, and spectral typed it as a T3 (blue) with a distance of 25.1 $\pm$ 0.32 pc. The other two, WISE 0004$-$2604 and WISE 0301$-$2319 were not subdwarf candidates. WISE 0004$-$2604 was a late-type candidate with an estimated spectral type, based on photometry of T0.5, and WISE 0301$-$2319 was a nearby late-type candidate with an estimated spectral type, based on photometry, of T0.5, and an estimated distance of 24--33 pc. We observed both of them with IRTF/SpeX, and typed WISE 0004$-$2604 as T2 (blue) with a distance of 25 $\pm$ 2.3 pc and WISE 0301$-$2319 as T1 (sl. blue) with a distance of 27 $\pm$ 2.3 pc.

If confirmed, these three objects would more than double the number of known early-type T subdwarfs. To date, only two early-type T subdwarfs are known: the sdT0 WISE  071121.36$-$573634.2 discovered by \cite{Kellogg2018} as part of the follow-up for the AllWISE2 motion survey \citep{Kirkpatrick2016}; and the sdT1.5 WISE 210529.08$-$623558.7, discovered by \cite{LuhmanandSheppard2014} as part of an analysis of high proper motion objects from the {\it WISE} survey. In addition, there are three published late-type T subdwarfs: the sdT5.5 HIP73786B, a common proper motion companion to the metal-poor K-star HIP73786 discovered by \cite{Murray2011} using data from the United Kingdom InfraRed Telescope (UKIRT) Infrared Deep Sky Survey (UKIDSS); the sdT6.5 ULAS J131610.28+075553.0 discovered by \cite{Burningham2014} in the UKIDSS Large Area Survey; and the sdT8 WISE J200520.38+542433.9, a companion to the sdM1.5 Wolf 1130, discovered by \cite{Mace2013} using photometry from 2MASS, WISE, and other telescopes. Although it was not initially designated as a T subdwarf, \cite{BBK2006} showed that the peculiar T6 dwarf 2MASS 0937+2931 has a sub solar metallicity and has a spectral morphology consistent with other T subdwarfs. In addition, \cite{Zhang2019} report 38 metal-poor T dwarfs, that show suppressed $K$-band flux in their spectra, which they believe might be T subdwarfs. All of these have spectral types of T5 or later.

As discussed above, there are gaps in our sequence of subdwarf spectral standards. We do have spectra of several subdwarf candidates at these missing spectral types including two candidate sdM4: WISE 1113+5010 and WISE 1710$-$1801; three new candidate sdL5: WISE 0328+0155, WISE 0413$-$2023, and WISE 2249$-$1627; and three new candidate T subdwarfs: WISE 0301$-$2319 (sdT1), WISE 0004$-$2604 (sdT2), and WISE 1019$-$3911 (sdT3). These objects could potentially be used to fill in these holes in the sequence. This is beyond the scope of this work but, in the future these spectra could aid in the construction of a more complete classification scheme for subdwarfs.

\begin{figure}
\centerline{\hbox{\includegraphics{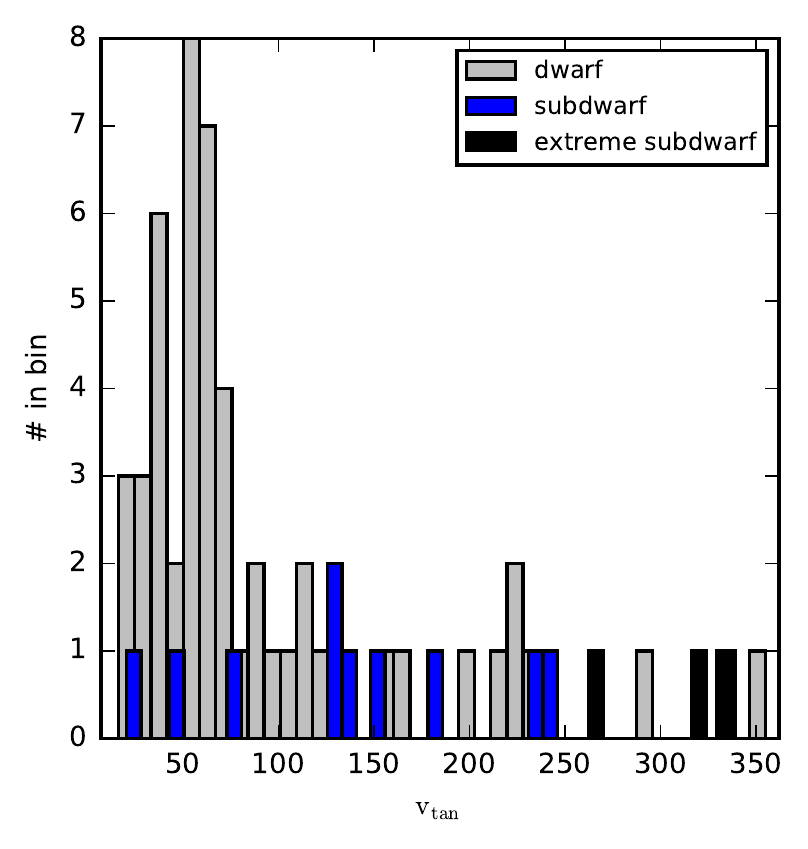}}}
\caption{Distribution of tangential velocities for our objects.  \label{fig:vtan}}
\end{figure}

We also calculated tangential velocities and their uncertainties for each of our objects, using a Monte Carlo approach to properly account for the uncertainties in the distances and proper motions. Normal distributions were constructed for $\mu_{\alpha}$, $\mu_{\delta}$, and distance, using their uncertainties. Values for each were randomly drawn from those distributions and used to calculate v$_\text{tan}$. This process was repeated 10,000 times, and the resulting distribution was fit to determine v$_\text{tan}$ and its uncertainty for each object. These values are reported in Table 5, and plotted in Figure \ref{fig:vtan}. In \cite{DupuyLiu2012}, they computed the membership probability as a function of tangential velocity for the thin disk, thick disk, and halo populations. Based on the results plotted in Figure 31 of that paper, we define the v$_\text{tan}$ values for these regions as follows: halo v$_\text{tan} \gtrsim 250$ km/s; thick disk 100 km/s $\lesssim$ v$_\text{tan}$ $\gtrsim$ 250 km/s; and thin disk v$_\text{tan} \lesssim 100$ km/s. All three of the extreme subdwarfs in our sample have v$_\text{tan} \gtrsim 250$ km/s, putting them in the halo, as expected of older, lower metallicity subdwarfs, which tend to be kinematically associated with the halo population. The dwarfs in our sample are likely clustered in the thin disk, though it is likely some are also in the thick disk. The subdwarfs in our sample are likely distributed throughout the thick and thin disk. Three of our dwarfs have tangential velocities that place them in the halo: WISE 0101+0336 (355.0 $\pm$ 30.1 km/s), WISE 0924+0723 (290 $\pm$ 26 km/s), and WISE 1113+5010 (460 $\pm$ 76 km/s). The velocity of  WISE 1113+5010 is approaching the escape velocity of the Galaxy, which is v$_\text{tan}$=$ 528  _{ -25}^{ +24}$ km/s at the Sun's position \citep{Deason2019}.

\begin{longrotatetable}
\begin{deluxetable*}{cccccccccc}
\tablecaption{Photometry, Proper Motions, Spectral Types, and Tangential Velocities of All Observed Objects}
\tabletypesize{\scriptsize}
\label{table:photometry}
\tablehead{
 \colhead{AllWISE} \vspace{-0.35cm} & \colhead{2MASS $J$} & \colhead{2MASS $H$} & \colhead{2MASS $K_{s}$} & \colhead{{\it WISE} $W$1} & \colhead{{\it WISE} $W$2} & \colhead{$\mu_{\alpha}$} & \colhead{$\mu_{\delta}$} & \colhead{v$_\text{tan}$} & \colhead{Spectral} \\
 \colhead{Designation} \vspace{-0.005cm} & \colhead{(mag)} & \colhead{(mag)} & \colhead{(mag)} & \colhead{(mag)} & \colhead{(mag)} & \colhead{(mas/yr)} & \colhead{(mas/yr)} & \colhead{km/s} & \colhead{Type} }
\startdata 
J000430.66$-$260402.3 & 16.487 $\pm$ 0.133 & 15.587 $\pm$ 0.129 & $>$15.523          & 15.211 $\pm$ 0.038 & 14.127 $\pm$ 0.044 & 11.9 $\pm$ 15.3     & $-$229.6 $\pm$ 13.8 &  27 $\pm$  2.9 & T2 (blue)     \\
J000458.47$-$133655.1 & 16.841 $\pm$ 0.171 & 16.120 $\pm$ 0.207 & $>$15.410          & 15.120 $\pm$ 0.037 & 14.457 $\pm$ 0.056 & 431.3 $\pm$ 21.8    & $-$37.4 $\pm$ 20.3  &  59 $\pm$  6.3 & T2           \\
J000536.63$-$263311.8 & 17.171 $\pm$ 0.225 & 15.849 $\pm$ 0.165 & 15.191 $\pm$ 0.154 & 14.924 $\pm$ 0.033 & 14.261 $\pm$ 0.047 & 384.0 $\pm$ 22.8    & 39.8 $\pm$ 20.5     &  56 $\pm$  5.8 & T0 (pec)      \\
J000856.39$-$281321.7 & 16.727 $\pm$ 0.137 & 15.664 $\pm$ 0.139 & 15.049 $\pm$ 0.131 & 14.119 $\pm$ 0.027 & 13.636 $\pm$ 0.037 & 284.3 $\pm$ 16.0    & $-$54.7 $\pm$ 13.6  &  40 $\pm$  4.0 & L8           \\
J010134.83+033616.0   & 15.418 $\pm$ 0.052 & 14.650 $\pm$ 0.066 & 14.300 $\pm$ 0.069 & 14.206 $\pm$ 0.029 & 13.941 $\pm$ 0.039 & 591.3 $\pm$ 10.0    & $-$557.8 $\pm$ 8.5  & 360 $\pm$ 30. & M7           \\
J010631.20$-$231415.1 & 17.338 $\pm$ 0.235 & 16.115 $\pm$ 0.204 & 15.683 $\pm$ 0.241 & 14.899 $\pm$ 0.033 & 14.417 $\pm$ 0.049 & $-$271.5 $\pm$ 24.8 & $-$201.6 $\pm$ 23.2 &  57 $\pm$  6.6 & L9           \\
J011049.18$+$192000.1 & 14.708 $\pm$ 0.032 & 14.142 $\pm$ 0.038 & 13.827 $\pm$ 0.05  & 13.467 $\pm$ 0.025 & 13.147 $\pm$ 0.03  & 451.9 $\pm$ 6.0     & 38.0 $\pm$ 6.0      & 130 $\pm$ 10. & M8           \\
J013525.38+020518.2   & 16.622 $\pm$ 0.129 & 15.481 $\pm$ 0.104 & 15.123 $\pm$ 0.117 & 14.283 $\pm$ 0.028 & 13.883 $\pm$ 0.04  & 102.9 $\pm$ 16.2    & $-$494.1 $\pm$ 15.2 &  60 $\pm$  5.4 & T0 (sl. red)   \\
J022721.93+235654.3   & 16.663 $\pm$ 0.135 & 15.647 $\pm$ 0.105 & 15.270 $\pm$ 0.155 & 14.304 $\pm$ 0.027 & 13.690 $\pm$ 0.035 & 310.0 $\pm$ 15.2    & $-$139.0 $\pm$ 13.6 &  44 $\pm$  4.2 & L9           \\
J030119.39$-$231921.1 & 16.635 $\pm$ 0.144 & 15.800 $\pm$ 0.158 & 15.579 $\pm$ 0.234 & 14.829 $\pm$ 0.03  & 14.036 $\pm$ 0.036 & 263.7 $\pm$ 27.9    & $-$141.4 $\pm$ 22.9 &  38 $\pm$  4.7 & T1 (sl. blue)  \\
J030919.70$-$501614.2\tablenotemark{a} & $\cdots$    & $\cdots$ & $\cdots$           & 16.465 $\pm$ 0.057 & 13.631 $\pm$ 0.031 & $\cdots$            & $\cdots$            &    $\cdots$      & T7           \\
J031627.79+265027.5   & 16.585 $\pm$ 0.149 & 15.592 $\pm$ 0.159 & $>$15.159          & 14.980 $\pm$ 0.035 & 13.934 $\pm$ 0.04  & 209.0 $\pm$ 22.5    & $-$15.6 $\pm$ 20.2  &  22 $\pm$  3.1 & T3           \\
J032309.12$-$590751.0 & 16.881 $\pm$ 0.189\tablenotemark{b} & $>$16.669          & $>$16.262          & 16.804 $\pm$ 0.065 & 14.529 $\pm$ 0.039 & 542.9 $\pm$ 24.0    & 476.7 $\pm$ 21.4    &  66 $\pm$  7.6 & T7           \\
J032838.73+015517.7   & 16.504 $\pm$ 0.172 & $>$16.598          & 15.202 $\pm$ 0.182 & 14.645 $\pm$ 0.031 & 14.327 $\pm$ 0.053 & 190.8 $\pm$ 22.7    & $-$233.2 $\pm$ 19.0 &  66 $\pm$  7.8 & L5 (blue)     \\
J033346.88+385152.6   & 16.073 $\pm$ 0.071 & 15.276 $\pm$ 0.082 & 15.000 $\pm$ 0.118 & 14.654 $\pm$ 0.03  & 14.350 $\pm$ 0.047 & 240.4 $\pm$ 9.7     & $-$324.3 $\pm$ 9.8  & 200 $\pm$ 17 & M8           \\
J034409.71+013641.5   & $>$17.024          & 15.949 $\pm$ 0.198 & 15.549 $\pm$ 0.232 & 14.647 $\pm$ 0.034 & 14.183 $\pm$ 0.05  & $-$72.0 $\pm$ 31.2  & $-$286.6 $\pm$ 30.3 &  51 $\pm$  7.2 & L8           \\
J034858.75$-$562017.8 & 16.652 $\pm$ 0.151 & $>$15.940          & $>$15.517          & 14.233 $\pm$ 0.028 & 13.919 $\pm$ 0.036 & 169.7 $\pm$ 18.0    & 206.6 $\pm$ 16.3    &  36 $\pm$  4.3 & T3           \\
J041353.96$-$202320.3 & 16.392 $\pm$ 0.111 & 15.444 $\pm$ 0.116 & 15.132 $\pm$ 0.174 & 14.233 $\pm$ 0.028 & 13.919 $\pm$ 0.036 & $-$41.8 $\pm$ 15.0  & $-$349.9 $\pm$ 14 &  20 $\pm$  2.7 & L5 (blue)     \\
J041743.13+241506.3   & 15.766 $\pm$ 0.069 & 15.654 $\pm$ 0.136 & 15.450 $\pm$ 0.167 & 14.520 $\pm$ 0.032 & 13.374 $\pm$ 0.035 & 403.6 $\pm$ 10.1    & $-$489.8 $\pm$ 10.2 &  35 $\pm$  3.2 & T6           \\
J053424.45+165255.0   & 15.445 $\pm$ 0.041 & 14.385 $\pm$ 0.037 & 13.572 $\pm$ 0.041 & 12.969 $\pm$ 0.024 & 12.575 $\pm$ 0.025 & $-$69.0 $\pm$ 6.4   & $-$76.4 $\pm$ 6.4   &  16 $\pm$  1.7 & L2 (pec)      \\
J054455.54+063940.3   & 14.039 $\pm$ 0.032 & 13.286 $\pm$ 0.028 & 12.795 $\pm$ 0.033 & 12.494 $\pm$ 0.024 & 12.265 $\pm$ 0.025 & 157.7 $\pm$ 12.8    & $-$329.2 $\pm$ 11.9 &  61 $\pm$  5.3 & M9           \\
J061429.77+383337.5   & 13.523 $\pm$ 0.024 & 12.748 $\pm$ 0.029 & 12.251 $\pm$ 0.02  & 11.848 $\pm$ 0.024 & 11.619 $\pm$ 0.022 & 84.9 $\pm$ 6.9      & $-$385.8 $\pm$ 6.8  &  50 $\pm$  4.2 & M9           \\
J062858.69+345249.2   & 15.957 $\pm$ 0.084 & 15.265 $\pm$ 0.089 & 14.706 $\pm$ 0.084 & 13.923 $\pm$ 0.027 & 13.615 $\pm$ 0.039 & 8.2 $\pm$ 7.4       & $-$286.5 $\pm$ 7.4  &  54 $\pm$  4.7 & L4           \\
J063552.52+514820.4   & $>$16.680          & 15.504 $\pm$ 0.144 & 15.416 $\pm$ 0.180 & 14.610 $\pm$ 0.031 & 14.294 $\pm$ 0.046 & $-$121.1 $\pm$ 18.9 & $-$243.2 $\pm$ 17.3 &  37 $\pm$  4.2 & T0           \\
J084254.56$-$061023.7 & 16.040 $\pm$ 0.076 & 15.680 $\pm$ 0.111 & $>$15.127          & 15.444 $\pm$ 0.041 & 14.086 $\pm$ 0.041 & $-$375.9 $\pm$ 14.9 & $-$45.0 $\pm$ 14.4  &  37 $\pm$  3.7 & T4           \\
J085039.11$-$022154.3 & 15.443 $\pm$ 0.044 & 14.504 $\pm$ 0.041 & 14.100 $\pm$ 0.059 & 13.408 $\pm$ 0.025 & 13.100 $\pm$ 0.028 & $-$392.0 $\pm$ 6.7  & $-$132.1 $\pm$ 6.7  &  47 $\pm$  6.6 & sdL7 (red)   \\
J085633.87$-$181546.6 & 15.828 $\pm$ 0.071 & 15.252 $\pm$ 0.094 & 14.473 $\pm$ 0.095 & 14.350 $\pm$ 0.029 & 14.178 $\pm$ 0.043 & 76.7 $\pm$ 8.3      & $-$251.5 $\pm$ 7.7  &  80 $\pm$  7.0 & L1           \\
J092453.76+072306.0   & 15.752 $\pm$ 0.083 & 15.272 $\pm$ 0.094 & 14.754 $\pm$ 0.112 & 14.722 $\pm$ 0.032 & 14.488 $\pm$ 0.054 & $-$248.4 $\pm$ 11.1 & $-$383.2 $\pm$ 10.7 & 290 $\pm$ 27 & M6           \\
J094812.21$-$290329.5 & 15.542 $\pm$ 0.056 & 15.019 $\pm$ 0.066 & 14.848 $\pm$ 0.122 & 14.332 $\pm$ 0.028 & 13.962 $\pm$ 0.03  & $-$370.3 $\pm$ 7.8  & $-$238.0 $\pm$ 8.0  & 150 $\pm$ 20 & sdL1         \\
J095230.79$-$282842.2 & 14.942 $\pm$ 0.043 & 14.451 $\pm$ 0.041 & 14.050 $\pm$ 0.058 & 13.934 $\pm$ 0.027 & 13.651 $\pm$ 0.034 & $-$572.2 $\pm$ 7.2  & 276.3 $\pm$ 7.2     & 340 $\pm$ 46 & esdM4        \\
J101944.62$-$391151.6 & 16.027 $\pm$ 0.096 & 15.766 $\pm$ 0.125 & 15.727 $\pm$ 0.267 & 15.645 $\pm$ 0.044 & 14.217 $\pm$ 0.042 & $-$472.2 $\pm$ 28.0 & 222.7 $\pm$ 26.4    &  58 $\pm$  5.9 & T3 (blue)     \\
J103534.63$-$071148.2 & 16.393 $\pm$ 0.094 & 15.843 $\pm$ 0.128 & 15.145 $\pm$ 0.141 & 14.381 $\pm$ 0.029 & 14.085 $\pm$ 0.045 & $-$375.5 $\pm$ 18.9 & $-$28.4 $\pm$ 15.8  &  80 $\pm$  11. & sdL7         \\
J111320.39+501010.5   & 15.506 $\pm$ 0.06  & 14.804 $\pm$ 0.086 & 14.898 $\pm$ 0.113 & 14.603 $\pm$ 0.029 & 14.411 $\pm$ 0.046 & $-$167.3 $\pm$ 10.8 & $-$313.2 $\pm$ 10.9 & 330$\pm$ 54 & M4 (blue)     \\
J112158.76+004412.3   & 15.961 $\pm$ 0.077 & 15.256 $\pm$ 0.075 & 14.877 $\pm$ 0.133 & 14.715 $\pm$ 0.033 & 14.297 $\pm$ 0.047 & $-$383.0 $\pm$ 9.5  & $-$133.0 $\pm$ 9.8  & 224 $\pm$ 20 & M7 (blue)     \\
J112859.45+511016.8   & 16.189 $\pm$ 0.069 & 15.110 $\pm$ 0.078 & 14.490 $\pm$ 0.069 & 13.944 $\pm$ 0.026 & 13.692 $\pm$ 0.032 & $-$117.7 $\pm$ 9.0  & $-$321.9 $\pm$ 9.1  &  74 $\pm$  6.4 & L3           \\
J120751.17+302808.9   & 15.253 $\pm$ 0.051 & 14.799 $\pm$ 0.071 & 14.467 $\pm$ 0.077 & 14.046 $\pm$ 0.027 & 13.713 $\pm$ 0.033 & 126.1 $\pm$ 9.1     & $-$241.1 $\pm$ 7.2  & 100 $\pm$  8.9 & M8           \\
J121231.97$-$050750.7 & 14.676 $\pm$ 0.032 & 14.201 $\pm$ 0.028 & 13.845 $\pm$ 0.049 & 13.663 $\pm$ 0.028 & 13.366 $\pm$ 0.033 & $-$474.2 $\pm$ 7.0  & $-$21.9 $\pm$ 7.1   & 160 $\pm$ 13 & M7 (sl. blue)  \\
J121914.75+081027.0   & 15.780 $\pm$ 0.085 & 15.076 $\pm$ 0.096 & 14.979 $\pm$ 0.148 & 14.796 $\pm$ 0.034 & 14.567 $\pm$ 0.059 & $-$279.9 $\pm$ 11.4 & $-$347.7 $\pm$ 11.1 & 250 $\pm$ 34 & sdM7         \\
J122042.20+620528.3   & 15.433 $\pm$ 0.054 & 14.830 $\pm$ 0.062 & 14.730 $\pm$ 0.086 & 14.487 $\pm$ 0.029 & 14.152 $\pm$ 0.034 & $-$467.9 $\pm$ 9.4  & $-$280.9 $\pm$ 8.7  & 230 $\pm$ 19 & sdM7         \\
J123513.87$-$045146.5 & 15.681 $\pm$ 0.072 & 15.195 $\pm$ 0.083 & 15.029 $\pm$ 0.142 & 14.808 $\pm$ 0.033 & 14.535 $\pm$ 0.057 & $-$230.0 $\pm$ 10.1 & $-$351.4 $\pm$ 9.9  & 317 $\pm$ 44 & esdM4        \\
J124516.66+601607.5   & 15.663 $\pm$ 0.058 & 15.297 $\pm$ 0.104 & 15.086 $\pm$ 0.116 & 14.711 $\pm$ 0.028 & 14.502 $\pm$ 0.043 & $-$294.7 $\pm$ 11.1 & $-$239.8 $\pm$ 9.7  & 190 $\pm$ 26 & sdM8.5       \\
J133520.09$-$070849.3 & 16.336 $\pm$ 0.097 & 15.365 $\pm$ 0.09  & 14.989 $\pm$ 0.134 & 14.932 $\pm$ 0.034 & 14.565 $\pm$ 0.056 & $-$367.5 $\pm$ 11.9 & 84.1 $\pm$ 12.0     & 230 $\pm$ 21 & M7           \\
J134359.71+634213.1   & 16.004 $\pm$ 0.085 & 15.202 $\pm$ 0.106 & 14.795 $\pm$ 0.096 & 14.476 $\pm$ 0.025 & 14.262 $\pm$ 0.033 & $-$254.0 $\pm$ 12.4 & 86.7 $\pm$ 8.3      & 130 $\pm$ 12 & M8           \\
J143942.79$-$110045.4 & 15.837 $\pm$ 0.086 & 15.365 $\pm$ 0.098 & 15.038 $\pm$ 0.145 & 14.586 $\pm$ 0.03  & 14.213 $\pm$ 0.043 & $-$252.3 $\pm$ 9.8  & $-$207.6 $\pm$ 9.3  & 130 $\pm$ 18 & sdL1         \\
J144056.64$-$222517.8 & 15.077 $\pm$ 0.05  & 14.688 $\pm$ 0.058 & 14.453 $\pm$ 0.087 & 14.108 $\pm$ 0.029 & 13.780 $\pm$ 0.042 & $-$239.5 $\pm$ 7.8  & $-$247.3 $\pm$ 7.9  & 130 $\pm$ 18 & sdM8.5       \\
J145645.54$-$103343.5 & 14.856 $\pm$ 0.049 & 14.197 $\pm$ 0.055 & 13.849 $\pm$ 0.044 & 13.467 $\pm$ 0.026 & 13.166 $\pm$ 0.031 & 29.2 $\pm$ 7.1      & $-$302.3 $\pm$ 7.1  &  87 $\pm$  7.6 & M8 (sl. blue)  \\
J145747.55$-$094719.3 & 15.331 $\pm$ 0.048 & 14.930 $\pm$ 0.063 & 14.688 $\pm$ 0.097 & 14.377 $\pm$ 0.029 & 14.205 $\pm$ 0.048 & $-$313.9 $\pm$ 8.6  & $-$250.9 $\pm$ 8.0  & 260 $\pm$ 36 & esdM4        \\
J155225.22+095155.5   & 15.923 $\pm$ 0.088 & 15.360 $\pm$ 0.082 & 15.164 $\pm$ 0.147 & 14.756 $\pm$ 0.03  & 14.554 $\pm$ 0.054 & $-$241.5 $\pm$ 12.1 & $-$291.8 $\pm$ 11.5 & 230 $\pm$ 33 & sdM7         \\
J165057.66$-$221616.8 & 12.218 $\pm$ 0.024 & 11.679 $\pm$ 0.027 & 11.332 $\pm$ 0.026 & 11.122 $\pm$ 0.023 & 10.929 $\pm$ 0.021 & $-$123.3 $\pm$ 5.9  & $-$266.1 $\pm$ 5.9  &  57 $\pm$  8.9 & M5           \\
J171059.52$-$180108.7 & 12.314 $\pm$ 0.027 & 11.800 $\pm$ 0.025 & $>$11.509          & 11.208 $\pm$ 0.024 & 11.027 $\pm$ 0.021 & $-$84.8 $\pm$ 7.5   & $-$365.1 $\pm$ 7.3  & 110 $\pm$ 22 & M4(blue)     \\
J171105.08$-$275531.7 & 12.760 $\pm$ 0.033 & 12.190 $\pm$ 0.036 & 11.853 $\pm$ 0.037 & 11.456 $\pm$ 0.024 & 11.315 $\pm$ 0.023 & $-$169.0 $\pm$ 6.2  & $-$373.0 $\pm$ 6.1  &  63 $\pm$  5.6 & M6           \\
J171454.88+064349.8   & 16.617 $\pm$ 0.132 & 15.467 $\pm$ 0.114 & 14.594 $\pm$ 0.089 & 14.066 $\pm$ 0.026 & 13.782 $\pm$ 0.036 & $-$83.2 $\pm$ 10.6  & $-$322.0 $\pm$ 10.6 &  88 $\pm$  8.0 & L2 (red)      \\
J173551.56$-$820900.3 & 16.393 $\pm$ 0.14  & $>$15.949          & $>$15.996          & 15.570 $\pm$ 0.036 & 13.723 $\pm$ 0.029 & $-$232.3 $\pm$ 16.3 & $-$253.4 $\pm$ 15.5 &  21 $\pm$  2.4 & T7           \\
J180839.55+070021.7   & 16.125 $\pm$ 0.104 & 15.731 $\pm$ 0.143 & 15.338 $\pm$ 0.169 & 14.821 $\pm$ 0.034 & 14.497 $\pm$ 0.055 & $-$235.1 $\pm$ 21.4 & $-$177.8 $\pm$ 19.8 & 110 $\pm$ 12 & L1 (blue)     \\
J182010.20+202125.8   & 15.188 $\pm$ 0.051 & 14.802 $\pm$ 0.071 & 14.606 $\pm$ 0.076 & 14.409 $\pm$ 0.03  & 14.174 $\pm$ 0.04  & $-$341.1 $\pm$ 8.5  & $-$45.4 $\pm$ 8.5   & 130 $\pm$ 19 & sdM8.5       \\
J183654.10$-$135926.2 & 12.997 $\pm$ 0.024 & 12.433 $\pm$ 0.023 & 12.031 $\pm$ 0.019 & 11.542 $\pm$ 0.028 & 11.480 $\pm$ 0.029 & $-$21.2 $\pm$ 6.5   & $-$368.2 $\pm$ 6.6  &  61 $\pm$  5.3 & M6           \\
J191011.03+563429.3   & 13.281 $\pm$ 0.027 & 12.654 $\pm$ 0.033 & 12.231 $\pm$ 0.026 & 11.825 $\pm$ 0.022 & 11.549 $\pm$ 0.021 & $-$364.6 $\pm$ 7.8  & 335.7 $\pm$ 6.9     &  68 $\pm$  5.7 & M8           \\
J201252.78+124633.3   & 12.040 $\pm$ 0.021 & 11.425 $\pm$ 0.021 & 11.035 $\pm$ 0.018 & 10.796 $\pm$ 0.023 & 10.596 $\pm$ 0.02  & 282.7 $\pm$ 6.0     & 151.7 $\pm$ 5.9     &  30 $\pm$  2.6 & M7 (sl. red)   \\
J211157.84$-$521111.3 & 16.563 $\pm$ 0.166 & 15.923 $\pm$ 0.212 & $>$15.252          & 15.371 $\pm$ 0.039 & 14.308 $\pm$ 0.043 & $-$227.2 $\pm$ 28.5 & 87.3 $\pm$ 26.7     &  30 $\pm$  4.4 & T3           \\
J215550.34$-$195428.4 & $>$16.978          & 15.971 $\pm$ 0.146 & 15.277 $\pm$ 0.142 & 14.552 $\pm$ 0.03  & 14.172 $\pm$ 0.044 & $-$34.0 $\pm$ 16.8  & $-$352.7 $\pm$ 16.1 &  66 $\pm$  6.7 & L7           \\
J221737.41$-$355242.7 & 14.874 $\pm$ 0.051 & 14.540 $\pm$ 0.068 & 14.236 $\pm$ 0.066 & 13.816 $\pm$ 0.025 & 13.573 $\pm$ 0.032 & 49.0 $\pm$ 7.1      & $-$304.9 $\pm$ 7.1  & 220 $\pm$ 35 & M5           \\
J223444.44$-$230916.1 & 15.262 $\pm$ 0.086 & 14.831 $\pm$ 0.11  & 14.082 $\pm$ 0.027 & 13.745 $\pm$ 0.037 & 16.121 $\pm$ 0.085 & 408.9 $\pm$ 24.8    & $-$26.8 $\pm$ 22.5  & 110 $\pm$ 12 & L5           \\
J224931.10$-$162759.6 & 17.328 $\pm$ 0.228 & 16.284 $\pm$ 0.24  & $>$14.679          & 14.843 $\pm$ 0.034 & 14.368 $\pm$ 0.053 & 374.5 $\pm$ 8.8     & 126.2 $\pm$ 8.7     &  69 $\pm$  6.0 & L7           \\
J230743.63+052037.3   & 14.741 $\pm$ 0.038 & 14.058 $\pm$ 0.027 & 13.763 $\pm$ 0.039 & 13.159 $\pm$ 0.024 & 13.032 $\pm$ 0.028 & $-$133.3 $\pm$ 8.5  & $-$107.5 $\pm$ 7.7  &  52 $\pm$  5.0 & M7           \\
J234404.85$-$250042.2 & 15.253 $\pm$ 0.056 & 14.634 $\pm$ 0.079 & 14.392 $\pm$ 0.083 & 13.881 $\pm$ 0.027 & 13.563 $\pm$ 0.034 & 342.7 $\pm$ 7.5     & $-$167.6 $\pm$ 6.8  & 150 $\pm$ 13 & M7           \\
\enddata
\tablenotetext{a}{WISE 030919.70$-$501614.2 does not show up in 2MASS, and therefore does not have a 2MASS-AllWISE proper motion, and also does not have a calculated v$_\text{tan}$. }
\tablenotetext{b}{The 2MASS $J$-band photometry for WISE 032309.12$-$590751.0 was taken from the 2MASS reject table.}
\end{deluxetable*}
\end{longrotatetable}

\acknowledgments
This paper includes data gathered from the 6.5 meter {\it Magellan} Telescopes located at Las Campanas Observatory, Chile, and is based in part on observations from Cerro Tololo Inter-American Observatory, National Optical Astronomy Observatory (NOAO Prop. ID 2016B-0003; PI: A. Schneider), which is operated by the Association of Universities for Research in Astronomy (AURA) under a cooperative agreement with the National Science Foundation. This publication makes use of data products from the {\it Wide-field Infrared Survey Explorer}, which is a joint project of the University of California, Los Angeles, and the Jet Propulsion Laboratory/California Institute of Technology, and NEOWISE which is a project of the Jet Propulsion Laboratory/California Institute of Technology. {\it WISE} and NEOWISE are funded by the National Aeronautics and Space Administration. This research has benefitted from the M, L, T and Y dwarf compendium hosted at DwarfArchives.org. This research made use of the SIMBAD database, operated at CDS, Strasbourg, France. 

\facilities{IRTF(SpeX), Magellan:Baade(FIRE), Blanco(ARCoIRIS)}

\software{Spextool (\citealt{Vacca2003Spex}; \citealt{Cushing2004Spex})}








\clearpage{}
\bibliographystyle{aasjournal}
\bibliography{/Users/jengreco/Jennifer/Toledo/CushingGroup/Papers/references}

\end{document}